\documentclass[traditabstract]{aa}
\usepackage{psfig}
\usepackage{graphicx,lscape,longtable}
\usepackage{rotating}

\begin{document}

\title{Accurate classification of 29 objects detected in the 39 months Palermo
Swift/BAT hard X-ray catalogue\thanks{
Based on observations obtained from the following observatories:  
Astronomical Observatory of Bologna in Loiano (Italy);
ESO-La Silla Observatory (Chile) under programme 083.D-0110; Observatorio Astron\'omico Nacional (San Pedro
M\'artir, Mexico), South African Astronomical Observatory, (South Africa).}}


\author{P. Parisi\inst{1,} \inst{4}, N. Masetti\inst{1}, E. Jim\'enez-Bail\'on\inst{2}, V. Chavushyan\inst{3}, E. Palazzi\inst{1}, R. Landi\inst{1}, A. Malizia\inst{1}
L. Bassani\inst{1}, A. Bazzano\inst{4}, A.J. Bird\inst{5}, P.A. Charles\inst{10}, G. Galaz\inst{6}, E. Mason\inst{7}, V.A. McBride\inst{10,} \inst{11}, D. Minniti\inst{6,} \inst{8}, 
L. Morelli\inst{9}, F. Schiavone\inst{1} and
P. Ubertini\inst{4}
}
\institute{
INAF -- Istituto di Astrofisica Spaziale e Fisica Cosmica di 
Bologna, Via Gobetti 101, I-40129 Bologna, Italy
\and 
Instituto de Astronom\'{\i}a, Universidad Nacional Aut\'onoma de M\'exico,
Apartado Postal 70-264, 04510 M\'exico D.F., M\'exico
\and
Instituto Nacional de Astrof\'{i}sica, \'Optica y Electr\'onica,
Apartado Postal 51-216, 72000 Puebla, M\'exico
\and
INAF -- Istituto di Astrofisica e Planetologia Spaziali di Roma, Via del Fosso del Cavaliere 100, Roma I-00133, Italy
\and
Physics \& Astronomy, University of Southampton, Southampton, Hampshire, SO171BJ, United Kingdom
\and
Departamento de Astronom\'{i}a y Astrof\'{i}sica, Pontificia Universidad 
Cat\'olica de Chile, Casilla 306, Santiago 22, Chile                     
\and
European Southern Observatory, Alonso de Cordova 3107, Vitacura,
Santiago, Chile
\and
Specola Vaticana, V-00120 Citt\`a del Vaticano
\and
Dipartimento di Astronomia, Universit\`a di Padova,
Vicolo dell'Osservatorio 3, I-35122 Padua, Italy
\and
South African Astronomical Observatory, P.O. Box 9, Observatory 7935,
South Africa
\and
University of Cape Town, Private Bag X3, Rondebosch 7701, South Africa
}

\offprints{P. Parisi (\texttt{pietro.parisi@iaps.inaf.it)}}
\date{Received 8 March 2012; accepted 15 June 2012}

\abstract{Through an optical campaign performed at 4 telescopes 
located in the northern and the southern hemispheres, plus 
archival data from two on-line sky surveys, we have obtained 
optical spectroscopy for 29 counterparts of unclassified or 
poorly studied hard X-ray emitting objects detected with 
{\it Swift}/BAT and listed in the 39 months Palermo  catalogue.
All these objects have also observations taken with {\it Swift}/XRT 
or {\it XMM}-EPIC which not only allow us to pinpoint their optical 
counterpart, but also to study their X-ray spectral properties (column density, power law photon index and F$_{2-10}$ keV flux).
We find that 28 sources in our sample are AGN; 7 are classified as type 1 while 21 are  of
type 2; the remaining object is a galactic cataclysmic variable. Among our type 1 AGN, we find 5 objects of intermediate Seyfert type (1.2-1.9) and 
one Narrow Line Seyfert 1 galaxy; for 4 out of 7 sources, we have been able to estimate the central black hole mass. Three of the type 2 AGN of our sample display optical features typical  of the LINER class and one is a likely Compton thick AGN.
All galaxies classified in this work are relatively nearby objects since their redshifts lie in the range 0.008-0.075; the only galactic object found lies at an estimated distance of 90 pc.
We have also investigated the optical versus X-ray emission ratio of the galaxies of our sample to test the AGN unified model. 
For them, 
we have also compared the X-ray absorption (due to gas) with the optical reddening (due to dust): we find that for most of our sources, specifically those of 
type 1.9-2.0 the former is higher than the latter confirming early results by Maiolino et al. (2001); this is possibly due to the properties of dust 
in the circumnuclear obscuring torus of the AGN.

}


\titlerunning{Classification of 29 objects in the Palermo
Swift/BAT catalogue}
\authorrunning{P. Parisi et al.}

\maketitle

\section{Introduction}
A critically important region of the astrophysical spectrum is the hard X-ray band, from 15 to 200 keV,
which is being recently explored in detail by two satellites,  {\it INTEGRAL} (Winkler et al. 2003) and {\it Swift} (Gehrels et al. 2004) which carry instruments 
like IBIS (Ubertini et al. 2003) and BAT (Barthelmy 2004) 
operating in the 20-200 keV band. These spacecrafts permit a study of the processes taking place in this observational window providing a deep look into the physics of hard X-ray sources.

These telescopes  operate in a complementary way, as the first concentrates on mapping the galactic plane, while the second mainly 
covers the high galactic latitude sky, so that together they provide the best sample of objects yet selected in the hard X-ray  domain. So far, 
both instruments have detected a 
large number of known and new objects, discovered new classes of sources and allowed finding and studying highly absorbed objects.
In particular, the nature of many of the objects detected above 20 keV by both satellites is often unknown, the sources are optically unclassified 
and their types can only be speculated on the basis of few available X-ray or radio observations. 

Optical follow up of these sources is therefore mandatory.
Specifically, the optical spectra can provide not only an accurate source classification, but also fundamental parameters which together with multiwaveband studies,
for example in the soft X-ray band, can provide information on these newly detected objects.

In this paper we focus on the X-ray and optical follow up work on a number of objects with unknown classification and/or redshift, reported in the 39 months 
Swift/BAT survey catalogue (Cusumano et al. 2010a). We note that the identifications of the present paper are also reported in the Palermo 54 months catalogue (Cusumano et al. 2010b) with preliminary classifications given by us via private communications.
Our aim is indeed to perform a systematic study of unidentified {\it Swift}/BAT objects starting with the 39 months surveys and continuing with the identifications of those of the 54 months catalogue (Parisi et al. in preparation).

This survey covers 90\% of the sky down to a flux limit of 2.5 $\times$ 10$^{-11}$ erg cm$^{-2}$ s$^{-1}$ and 
50\% of the sky down to a flux limit of 1.8 $\times$ 10$^{-11}$ erg cm$^{-2}$ s$^{-1}$ in the 14-150 keV band. It lists 754 sources, 
of which 69\% are extragalactic, 27\% are galactic, 
and 4\% are of unknown type.

Within this BAT survey, we have selected a sample of 29 objects either without optical identification,
or not well studied, or without published optical spectra.
For all  these sources we first performed the X-ray data analysis to reduce the source positional uncertainty from arcmin to arcsec-sized radii and derive 
information on the main spectral parameters (photon index, column density and 2-10 keV flux). Then, within the reduced X-ray error boxes, we identified the putative optical counterpart
 to the BAT object  and performed 
optical spectroscopic follow up work. Following the method applied by Masetti et al. (2004, 2006a,b, 2008, 2009, 2010, 2012) and Parisi et al. (2009) 
we determined the nature of all selected objects and discussed their properties.
A preliminary classification of these sources is given in the Palermo 54 months BAT catalogue (Cusumano et al. 2010b), while here we publish for the first time the optical spectra and detailed optical
information (see Tab. \ref{cvo}, \ref{agn1} and \ref{agn2}).

We also checked  for the presence of peculiar sources, such as Compton thick AGN, absorbed Seyfert 1 and unabsorbed Seyfert 2, using the diagnostic method of
Malizia et al. (2007), and finally we used the plot of Maiolino et al. (2001) to verify a possible mismatch between the X-ray gas absorption and 
the optical dust reddening.

The paper is structured as follow:  
in Sect. 2 we report information on the soft X-ray data analysis; in Sect. 3 we give a description of the optical observations, the telescope employed, 
and provide information on the data reduction method used. 
Sect. 4 reports and discusses the main optical results  (line fluxes, distances, galactic and local extinction, 
central black hole masses etc.). In Sect. 5 the X-ray and the optical results are compared in view of the object classification and gas versus dust absorption. 
In Sect. 6, we summarize the main
conclusions of our work. 

\section{X-ray data analysis}
In this section we provide general information about the X-ray data analysis performed for the 29 objects of our sample, in order to obtain indications on the X-ray counterpart of the BAT object, provide its position with arcsec accuracy (see Table \ref{log}) and finally study
its spectral properties in the 2-10 keV band (see Table \ref{fit}).

For 24 out of 29 objects we use X-ray data acquired with the X-ray Telescope (XRT, 
0.3--10 keV, Burrows et al. 2004) on board the {\it Swift} satellite.
The XRT data reduction was performed using the XRTDAS standard data pipeline package ({\sc xrtpipeline} 
v. 0.12.6), in order to produce screened event files. All data were extracted only in the Photon Counting 
(PC) mode (Hill et al. 2004), adopting the standard grade filtering (0--12 for PC) according to the XRT 
nomenclature.  Depending on the source nature (bright or dim)  we either used the longest exposure or added multiple observations 
to enhance the signal to noise ratio.
For each BAT detection we then analysed, with {\sc XIMAGE} v. 4.5.1, the 3-10 keV  image of interest (single or added over more XRT pointings) to search for 
sources detected (at a confidence level $>$ 3$\sigma$) within the 90$\%$ {\it Swift}/BAT error circles; the 3-10 keV image choice assures 
that we are selecting the hardest sources and hence the most likely counterparts to the BAT objects.     
We estimated the X-ray positions and relative uncertainties  using the task {\sc xrtcentroid v.0.2.9}.

For  5  sources, we use instead X-ray data acquired with the pn X-ray CCD camera on
the EPIC instrument on-board the \emph{XMM-Newton} spacecraft (Str\H{u}der et al. 2001) in order to have a better estimate of the spectral properties of those objects with low quality XRT data.
These data were processed using the Standard Analysis Software (SAS) version 9.0.0
employing the latest available calibration files.
Only patterns corresponding to single and double events (PATTERN$\leq$4) were taken
into account and the standard selection filter FLAG=0 was applied.
For each source we analysed a single observation, the longest in terms of exposure, in order to have better statistics,  (see Table \ref{id} for the observation IDs). In each case we searched the \emph{XMM-Newton} EPIC-pn images for X-ray sources 
which fall inside the \emph{Swift}/BAT error circles; again to avoid false associations with dim and soft objects,  we only inspected higher energy images 
($>$4 keV in this case). \\
In 28 out of 29 objects analysed, a single object was detected within the BAT 90$\%$ positional uncertainty;  the exception is  PBC J0041.6+2534 for which
we find one soft X-ray source located just outside  the 90$\%$ error circle and none within it. We nevertheless consider it as the likely counterpart of the BAT source.

Next, we analysed the X-ray spectrum of each object:
events for spectral analysis were extracted within a circular region of radius 
20$^{\prime \prime}$, centered on the source position for XRT\footnote{This region encloses about 90$\%$ of the PSF at 1.5 
keV (see Moretti et al. 2004)} and  choosing instead the radius corresponding to the 
highest Signal-to-Noise ratio for EPIC-pn. The background was taken from empty regions close to the X-ray source 
of interest, using circular regions with different radii for XRT data, in order to ensure an evenly sampled background. In the case of EPIC-pn data
we adopted instead an 80$^{\prime\prime}$ radius for all the 5 sources analysed. 
XRT spectra  were then extracted from the corresponding event files using the {\sc XSELECT v.2.4} 
software and binned using {\sc grppha} in an appropriate way, so that the $\chi^{2}$ statistic could be 
applied.  We used version v.011 of the response matrices and created the relative ancillary response file 
\textit{arf} using the task {\sc xrtmkarf v.0.5.6}. 
\begin{landscape}
\begin{table}[th!]
\caption[]{Log of the spectroscopic observations presented in this paper
(see text for details). Optical source coordinates
are extracted from the 2MASS catalog and have an accuracy better than 0$\farcs$1, soft X-ray coordinates are extracted from XRT and \emph{XMM-Newton} observations.}
\label{log}
\scriptsize
\begin{center}
\resizebox{25cm}{!}{
\begin{tabular}{lllcllccclr}
\noalign{\smallskip}
\hline
\hline
\multicolumn{1}{c}{{\it (1)}} & \multicolumn{1}{c}{{\it (2)}} & \multicolumn{1}{c}{{\it (3)}} & \multicolumn{1}{c}{{\it (4)}} &  \multicolumn{1}{c}{{\it (5)}}&\multicolumn{1}{c}{{\it (6)}}&
{\it (7)} & {\it (8)} & {\it (9)} & \multicolumn{1}{c}{{\it (10)}}& \multicolumn{1}{c}{{\it (11)}}\\
\multicolumn{1}{c}{Object} & \multicolumn{1}{c}{RA X-ray} & \multicolumn{1}{c}{Dec X-ray} &\multicolumn{1}{c}{Error radius} & \multicolumn{1}{c}{RA Opt} &\multicolumn{1}{c}{Dec Opt}&\multicolumn{1}{c}{Telescope+instrument} & $\lambda$ range & Disp. & \multicolumn{1}{c}{UT Date \& Time}  & Exposure \\
& \multicolumn{1}{c}{(J2000)} & \multicolumn{1}{c}{(J2000)} & (arcsec) &\multicolumn{1}{c}{{\it (arcsec)}}  &\multicolumn{1}{c}{{\it (arcsec)}} & & (\AA) & (\AA/pix) & 
\multicolumn{1}{c}{at mid-exposure} & time (s)  \\
\noalign{\smallskip}
\hline
\noalign{\smallskip}
PBC J0041.6+2534$^{+*}$   &00 41 27.98 &+25 29 57.78     & 4.59   &00 41 28.01  &+25 29 57.6     & SPM 2.1m+B\&C Spc. &3450-7650& 4.0 & 28 Jan 2009, 03:18 & 2$\times$1800  \\
PBC J0100.6$-$4752        &01 00 34.54 &$-$47 52 05.71 & 3.75   &01 00 34.90 &$-$47 52 03.3    & AAT+6dF & 3900-7600 & 1.6 & 30 Oct 2010, 14:34 &   1200+600 \\
PBC J0122.3+5004           &01 22 34.35 &+50 03 12.24  & 4.59   & 01 22 34.42 &+50 03 18.0	   & SPM 2.1m+B\&C Spc. & 3450-7650  & 4.0 & 29 Jan 2009, 03:48 & 2$\times$1800    \\
PBC J0140.4$-$5320      &01 40 26.88  &$-$53 19 37.28  & 3.62   & 01 40 26.78 &$-$53 19 39.2   & Radcliffe+Gr. Spec. & 3850-7200  & 2.3 & 06 Aug 2009, 02:46 & 2$\times$900  \\
PBC J0248.9+2627           &02 48 59.17 &+26 30 38.27  & 4.36   & 02 48 59.37 &+26 30 39.1	   & SPM 2.1m+B\&C Spc. &3450-7650& 4.0 & 02 Dec 2008, 07:28 & 2$\times$1800	   \\
PBC J0353.5+3713           &03 53 42.28 &+37 14 06.06  & 3.90   & 03 53 42.46 &+37 14 07.7	   &  SPM 2.1m+B\&C Spc. &3450-7650& 4.0 & 28 Jan 2009, 05:33 & 2$\times$1800	\\
PBC J0356.9$-$4040      &03 56 56.40 & $-$40 41 45.50  & 3.65   & 03 56 56.55 &$-$40 41 45.3   & AAT+6dF & 3900-7600 & 1.6 & 19 Dec 2003, 10:23 & 1200+600\\
PBC J0503.0+2300          &05 02 58.10 & +22 59 50.41  & 3.53   & 05 02 58.22 &+22 59 52.0	   & SPM 2.1m+B\&C Spc. &3450-7650& 4.0 & 28 Jan 2009, 07:58 & 2$\times$1800 \\
PBC J0543.6$-$2738      &05 43 33.07 &$-$27 39 07.61  & 3.8    & 05 43 32.92&$-$27 39 05.3    &  SPM 2.1m+B\&C Spc. &3450-7650& 4.0  & 02 Dec 2008, 09:12 & 2$\times$1800  \\
PBC J0544.3+5905          &05 44 22.29 &+59 07 34.51   &3.68    & 05 44 22.57 &+59 07 36.1	  & SPM 2.1m+B\&C Spc. &3450-7650& 4.0 & 03 Dec 2008, 08:47 & 2$\times$1800  \\
PBC J0623.8$-$3212     &06 23 46.37 &$-$32 12 59.75    & 4.88   & 06 23 46.41 &$-$32 12 59.9	&  AAT+6dF & 3900-7600 & 1.6  & 07 Jan 2002, 12:30 &1200+600\\
PBC J0641.3+3251           &06 41 18.00 &+32 49 32.28  & 3.71   & 06 41 18.06 &+32 49 31.3	  & Cassini+BFOSC & 3500-8000 & 4.0 & 09 Dec 2009, 20:43 & 2$\times$1800  \\
PBC J0759.9+2324          &07 59 53.62 &+23 23 23.29   & 3.79   & 07 59 53.47 &+23 23 24.1	  & SDSS+CCD Spc. & 3800-9200 & 1.0 &09 Mar 2003, 02:39  & 2220  \\
PBC J0814.4+0421          &08 14 25.22 &+04 20 30.51    & 4.32   & 08 14 25.29 &+04 20 32.4	   &  SPM 2.1m+B\&C Spc. &3450-7650& 4.0 & 28 Jan 2009, 09:40 & 2$\times$1800	    \\
PBC J0826.3$-$7033     &08 26 23.19&$-$70 31 42.18     & 3.75   & 08 26 23.50 &$-$70 31 43.1   & NTT+EFOSC2    & 3650-9300  & 5.5 & 30 May 2009, 23:55 &  900 \\ 
PBC J0919.9+3712$^*$   &09 19 57.90 &+37 11 25.61      &3.0     & 09 19 58.02 &+37 11 27.7		 & SDSS+CCD Spc.&3800-9200&1.0&  29 Nov 2003, 10:11  & 2370 \\
PBC J0954.8+3724           &09 54 39.51 &+37 24 30.84  & 4.64   & 09 54 39.43 &+37 24 30.8	   & SDSS+CCD Spc.&3800-9200&1.0& 25 Dec 2003, 07:23  & 2250 \\
PBC J1246.5+5432$^*$      &12 46 40.01 &+54 31 59.77   & 5.08   & 12 46 39.82 &+54 32 03.0	& SPM 2.1m+B\&C Spc. &3450-7650& 4.0 & 28 Jan 2009, 11:49 & 2$\times$1800  \\
PBC J1335.8+0301$^*$     &13 35 48.44 &+02 59 54.41    & 4.28   & 13 35 48.25 &+02 59 55.6	&  SDSS+CCD Spc.&3800-9200&1.0& 23 Apr 2001, 06:07  & 7043 \\
PBC J1344.2+1934          &13 44 15.69 &+19 33 59.61    & 6.41   &13 44 15.80 &+19 33 57.9	    & SPM 2.1m+B\&C Spc. &3450-7650& 4.0 & 30 Jan 2009, 04:45 & 2$\times$1800  \\
PBC J1345.4+4141          &13 45 19.33 &+41 42 45.15   & 4.28   & 13 45 19.13 &+41 42 44.4	    & SDSS+CCD Spc.&3800-9200&1.0& 15 Apr 2004, 08:41  & 2400 \\
PBC J1439.0+1413          &14 39 11.87 &+14 15 18.46   & 3.91   & 14 39 11.86 &+14 15 21.5	    & SPM 2.1m+B\&C Spc. &3450-7650& 4.0 & 01 Feb 2009, 12:52  & 2$\times$1800  \\
PBC J1453.0+2553          &14 53 07.66 &+25 54 33.99   & 3.55   & 14 53 07.91 &+25 54 33.2	    & SPM 2.1m+B\&C Spc. &3450-7650& 4.0 & 31 Jan 2009, 11:48  &1800+1200  \\
PBC J1506.6+0349          &15 06 43.95 &+03 51 45.43   & 3.68   & 15 06 44.12 &+03 51 44.4	    & SPM 2.1m+B\&C Spc. &3450-7650& 4.0 & 01 Feb 2009, 10:48  &2$\times$1800  \\
PBC J1546.5+6931          &15 46 23.85 &+69 29 10.74   & 4.55   & 15 46 24.24 &+69 29 10.2	    & Cassini+BFOSC & 3500-8000 & 4.0 & 18 May 2009, 22:06 & 2$\times$1800  \\
PBC J1620.3+8101          &16 19 19.94 & +81 02 46.30  &  3.74   & 16 19 19.31 & +81 02 47.3  & Cassini+BFOSC  & 3500-8000 & 4.0 & 15 May 2009, 23:34 &1800 \\
PBC J2148.2$-$3455       &21 48 19.21 &$-$34 57 02.77  & 5.24   & 21 48 19.48 &$-$34 57 04.7		 & AAT+6dF & 3900-7600 & 1.6  &07 Jul 2003, 16:40  &1200+600 \\
PBC J2333.9$-$2343      &23 33 55.25 &$-$23 43 41.76   & 3.54   & 23 33 55.20 &$-$23 43 40.6		 & SPM 2.1m+B\&C Spc. &3450-7650& 4.0 & 18 Sep 2009, 07:48  &2$\times$1800  \\
PBC J2341.9+3036$^*$    &23 41 55.25 &+30 34 53.97     & 4.45     & 23 41 55.45 &+30 34 54.2	  & SPM 2.1m+B\&C Spc. &3450-7650& 4.0 & 02 Dec 2008, 03:10 &2$\times$1800  \\
\noalign{\smallskip}
\hline
\hline
\noalign{\smallskip} 
\multicolumn{11}{l}{Note: if not indicated otherwise, source optical coordinates were extracted from the 2MASS 
catalog and have an accuracy better than 0$\farcs$1.}\\
\multicolumn{11}{l}{$^*$ The reported X-ray coordinates are obtained from XMM/EPIC data. Our positions totally agree with those in the 2XMM catalogue (Watson et al. 2009).} \\ 
\multicolumn{11}{l}{$^+$ This source is outside  BAT 90\% error box, but inside the 99\% one.} \\
\noalign{\smallskip}
\noalign{\smallskip}
\end{tabular}}
\end{center}
\end{table}
\end{landscape}

For XMM spectra,
ancillary response  matrices (ARFs) and  detector response matrices
(RMFs) were generated using the \emph{XMM-Newton} SAS tasks \emph{arfgen} and
\emph{rmfgen}, respectively. In general, the spectral channels were rebinned in order to achieve a 
minimum of 20 counts in each bin (see however Table \ref{fit}).

The energy band used for the spectral analysis, 
performed with {\sc XSPEC} v.12.6.0, depends on the statistical quality of the data and 
typically ranges from 0.3 to $\sim$6 keV for XRT and from 0.5 to 12 keV for EPIC-pn.

In the first instance we adopted, as our
basic model, a simple power law passing through Galactic (Dickey \& Lockman 1990) and (when required) intrinsic absorption. 

\begin{table}[h!]
\caption[]{Observations IDs of the \emph{XMM-Newton} sources presented in this paper.}
\label{id}
\begin{center}
\begin{tabular}{lc}
\noalign{\smallskip}
\hline
\hline
\noalign{\smallskip}
\multicolumn{1}{c}{Object} & Observation IDs\\
\hline
\noalign{\smallskip}

PBC J0041.6+2534   & 0153030101  \\
PBC J0919.9+3712 & 0149010201  \\
PBC J1246.9+5432 & 0554500101 \\
PBC J1335.8+0301 & 0601781201    \\
PBC J2341.9+3036 & 0554500501  \\
\noalign{\smallskip} 
\hline
\hline
\end{tabular}
\end{center}
\end{table}

If this baseline model was not sufficient to fit the data, we introduced extra spectral components (i.e. Black body, second power law or gaussian line) as required according to the F-test statistics. 

Note that in some cases due to the limited statistical quality of the XRT data
we have fixed the photon index to 1.8 (canonical value for AGN; e.g., Mushotzky et al. 1993) in order to estimate the intrinsic column density and/or the X-ray flux. 

We are aware that in a few cases the statistical quality of the data is such that the parameter values obtained must be taken with caution; 
nevertheless they provide some indications on the source properties, i.e. if a source is likely absorbed or not.

The results of the X-ray spectral analysis are reported in Table \ref{fit}, where we list for each object the galactic column density, 
the power law photon index, the column density in excess to the galactic value, the reduced  
$\chi^{2}$ of the best-fit model, the 2--10 keV flux and the 20--100 keV flux\footnote{Hard X-ray fluxes have been extrapolated from 15-150 keV fluxes 
assuming a power law with $\Gamma$ = 2.02 (see Molina et al. 2012 in preparation)}; extra spectral parameters, if required, are reported 
in the notes at the end of the Table.
Here and in the following all quoted errors correspond to a 90$\%$ confidence level for a single X-ray 
parameter of interest ($\Delta\chi^{2}=2.71$).

Note that for a limited number of our sources X-ray data have  already been published: Winter et al. (2009) discuss the spectra 
of PBC J0641.3+3251 and PBC J0356.9-4040 while Noguchi et al. (2009) analysed the \emph{XMM-Newton} of PBC J0919.9+3712.
Finally PBC J2148.2-3455 has been extensively studied at X-ray energies using various instruments such as \emph{XMM-Newton} and Chandra 
(see for example Gonz\'alez-Mart\'in et al. 2009; Levenson et al. 2005), but no XRT data have been presented before.
Overall we find good agreement with these previous studies.

\begin{table*}
\begin{center}
\caption[]{Main results obtained from the analysis of the X-ray spectra of all AGN present in the
sample.}
\label{fit}
\scriptsize
\begin{tabular}{lcccccc}
\noalign{\smallskip}
\hline
\hline
\noalign{\smallskip}
\multicolumn{1}{c}{Source} & \multicolumn{1}{c}{N$_{Hgal}$}  & $\Gamma^{+}$ & N$_H$  &  $\chi^2/\nu$   & F$_{(2-10)keV}$ &F$_{(20-100)keV}$ \\
\noalign{\smallskip}
&$\times 10^{22}$cm$^{-2}$ & &$\times 10^{22}$cm$^{-2}$ & &$\times 10^{-11}$erg s$^{-1}$ cm$^{-2}$&$\times 10^{-11}$erg s$^{-1}$ cm$^{-2}$ \\
\noalign{\smallskip}
\hline
\noalign{\smallskip}

\bf PBC J0041.6+2534$^{A*}$  &\bf  0.038 & \bf2.56$^{+0.31}_{-0.21}$& \bf21.2$^{+11.4}_{-7.8}$ & \bf 6.2/7& \bf 0.02 &\bf 1.0  \\

& & & & & & \\

PBC J0100.6$-$4752$^A$ & 0.019 & 1.67$^{+0.37}_{-0.54}$ & 4.5$^{+1.8}_{-1.6}$ & 4.3/11 & 0.28 & 0.6 \\

& & & & & & \\

PBC J0122.3+5004$^A$ & 0.136 &2.22$^{+0.70}_{-0.60}$& 44$^{+129}_{-35}$  & 1.9/3 & 0.09 & 0.6 \\

& & & & & & \\

PBC J0140.4$-$5320$^A$ &0.026  & 1.58$^{+0.25}_{-0.23}$ & 1.2$^{+0.4}_{-0.3}$& 28.7/26& 0.50 & 0.7 \\

& & & & & & \\

PBC J0248.9+2627 & 0.103 & [1.8] &41.2$^{+20.8}_{-14.2}$ & 3.6/3 & 0.40 & 1.2 \\

& & & & & & \\

PBC J0353.5+3713$^A$ & 0.168 &1.71$^{+0.59}_{-0.55}$ & 3.7$^{+2.0}_{-1.7}$ & 13.2/14  & 0.35 & 0.8 \\

& & & & & &\\

PBC J0356.9$-$4040$^A$ & 0.019 & 1.83$^{+0.16}_{-0.31}$  & 4.0$^{+0.9}_{-0.9}$ &25.9/38 & 0.74  & 1.1 \\
& & & & & & \\

PBC J0503.0+2300 &0.226 & 2.11$^{+0.08}_{-0.08}$ & 0.11$^{+0.03}_{-0.02}$  & 148.6/149 & 1.25& 1.1\\

& & & & & & \\

PBC J0543.6$-2738$$^A$  &0.022  & 2.38$^{+0.46}_{-0.50}$& 3.6$^{+1.9}_{-1.2}$ &8.3/8 & 0.47& 1.1 \\

& & & & & & \\

PBC J0544.3+5905    &0.158 &   1.22$^{+0.44}_{-0.29}$ & 1.5$^{+0.7}_{-0.4}$ &8.3/18 & 0.47 & 1.1  \\

& & & & & & \\

PBC J0623.8$-$3212$^B$   &0.038 & [1.8]& 43.9$^{+15.8}_{-8.7}$  & 3.3/2 & 0.07 & 1.5\\

& & & & & & \\

PBC J0641.3+3251$^A$ & 0.159& 1.72$^{+0.32}_{-0.21}$& 12.2$^{+4.2}_{-3.7}$ &14.5/16 &  0.29& 1.0 \\

& & & & & & \\

PBC J0759.9+2324 & 0.047  & 1.57$^{+0.56}_{-0.49}$ & 1.6$^{+0.9}_{-0.5}$ &  9/10 & 0.44& 1.3\\

& & & & & & \\

PBC J0814.4+0421        & 0.032 &  1.2$^{+0.80}_{-0.65}$ &  5.6$^{+2.9}_{-2.0}$ & 9.3/15  & 0.50 & 1.1\\



& & & & & & \\
\bf PBC J0919.9+3712$^{C1}$ &\bf 0.011  &\bf 1.66$^{+0.07}_{-0.07}$  &\bf 7.2$^{+0.4}_{-0.4}$ &\bf 326.9/328 &\bf 0.36 &\bf 1.2 \\

& & & & & & \\

PBC J0954.8+3724$^*$ & 0.014 & [1.8] &21.4$^{+13.8}_{-8.3}$ & 2.4/2 & 0.15& 0.4 \\

& & & & & & \\

\bf PBC J1246.9+5432$^D$ &\bf 0.014 & \bf 0.88$^{+0.13}_{-0.12}$ &\bf 23.6$^{+8.0}_{-9.6}$  & \bf15/22 &\bf 0.09 &\bf 1.5 \\

& & & & & & \\

\bf PBC J1335.8+0301$^A$&\bf 0.019 &\bf 1.58$^{+0.06}_{-0.04}$ &\bf 2.3$^{+0.1}_{-0.1}$ &\bf 313.3/299 &\bf 0.64 &\bf 1.1  \\

& & & & & &\\

PBC J1344.2+1934   &0.017  &[1.8] & 38$^{+56}_{-20}$ & 2/4 & 0.10 &0.9\\

& & & & & & \\

PBC J1345.4+4141$^A$ & 0.009 &1.38$^{+0.27}_{-0.25}$ & 0.8$^{+0.4}_{-0.3}$& 10.4/25 & 0.74 & 0.7 \\

& & & & & &\\

PBC J1439.0+1413$^A$ & 0.014  &[1.8] & 3.4$^{+1.1}_{-1.0}$ & 12/8 & 0.32 &0.5 \\

& & & & & & \\

PBC J1453.0+2553   & 0.033 & 1.72$^{+0.05}_{-0.05}$ & -- & 73.8/87 & 1.06&1.0\\

& & & & & & \\

PBC J1506.6+0349   & 0.037 &  1.36$^{+0.35}_{-0.28}$& 1.1$^{+0.4}_{-0.3}$ & 14/19 & 0.42 &0.7 \\

& & & & & &\\

PBC J1546.5+6931$^*$ & 0.031  & [1.8] &  16$^{+26}_{-8.0}$   & 3.3/2& 0.08 &0.5\\

& & & & & & \\

\hline
\hline
\end{tabular}
\end{center}
\end{table*} 

\begin{table*}
\setcounter{table}{2}
\caption{-- \emph{continued}}
\scriptsize
\begin{center}
\begin{tabular}{lcccccc}
\hline
\hline
\noalign{\smallskip}
\multicolumn{1}{c}{Source} & \multicolumn{1}{c}{N$_{Hgal}$}  & $\Gamma^{+}$ & N$_H$  &  $\chi^2/\nu$   & F$_{(2-10)keV}$ &F$_{(20-100)keV}$ \\
\noalign{\smallskip}
&$\times 10^{22}$cm$^{-2}$ & &$\times 10^{22}$cm$^{-2}$ & &$\times 10^{-11}$erg s$^{-1}$ cm$^{-2}$&$\times 10^{-11}$erg s$^{-1}$ cm$^{-2}$ \\
\noalign{\smallskip}
\hline
\noalign{\smallskip}
PBC J1620.3+8101 & 0.046 & [1.8]& 10.1$^{+2.1}_{-1.6}$ & 9.6/13& 0.54 & 0.7 \\

& & & & & & \\

PBC J2148.2$-$3455 &0.019  &[1.8] &  --&6/11 & 0.04 &0.9 \\

& & & & & & \\

PBC J2333.9$-$2343 & 0.016& 1.67$^{+0.04}_{-0.04}$ & -- & 97/104 & 0.80 &0.7  \\

& & & & & & \\

\bf PBC J2341.9+3036$^{C2}$&\bf 0.058  &\bf 2.02$^{+0.16}_{-0.15}$ &\bf 56.5$^{+15.5}_{-10.4}$ &\bf 25/28 &\bf 0.11 &\bf 1.0 \\

& & & & & &\\
\noalign{\smallskip} 
\hline
\hline
\multicolumn{7}{l}{$^{+}$ The square brackets in the $\Gamma$ column indicate that we used a fixed value}\\
\multicolumn{7}{l}{$^{*}$ Grouping of 10 instead of 20  applied due to the poor statistical quality of the data}\\
\multicolumn{7}{l}{$^{A}$  Best fit model requires an extra power law component, having the same photon index of the primary absorbed power law }\\
\multicolumn{7}{l}{and passing only through the galactic column density (wa$_{gal}$*(po+wa*po) in xspec terminology)}\\
\multicolumn{7}{l}{$^{B}$ Best fit model requires  an extra black body component with kT=0.16$^{+0.04}_{-0.03}$  keV 
(wa$_{gal}$*(bb+wa*po) in xspec terminology)}\\
\multicolumn{7}{l}{$^{C1}$ Best-fit model requires an extra power law component, having the same photon index of the primary absorbed power law and}\\
\multicolumn{7}{l}{passing only through the galactic column density plus a narrow line at $E = 6.36^{+0.03}_{-0.03}$ keV with  EW= $194^{+20}_{-42}$ eV  }\\
\multicolumn{7}{l}{(wa$_{gal}$*(po+wa*(po+ga)) in xspec terminology)} \\
\multicolumn{7}{l}{$^{D}$ Best-fit model requires a black body component with a $kT = 0.28^{+0.04}_{-0.03}$ keV and two narrow lines at $E = 6.29^{+0.03}_{-0.03}$ keV}\\ 
\multicolumn{7}{l}{and $6.79^{+0.11}_{-0.10}$ keV with an EW  of $600^{+182}_{-174}$ eV and $378^{+168}_{-167}$ eV respectively
(wa$_{gal}$*(bb+wa*(po+ga+ga)) in xspec terminology)} \\
\multicolumn{7}{l}{$^{C2}$ Best-fit model requires a second power law component, having the same photon index of the primary absorbed power law }\\
\multicolumn{7}{l}{and passing only through the Galactic column density plus  a narrow line  with $E = 6.25^{+0.05}_{-0.06}$ and $365^{+176}_{-147}$ eV} \\ 
\multicolumn{7}{l}{(wa$_{gal}$*(po+wa*(po+ga)) in xspec terminology)} \\
\multicolumn{7}{l}{\bf In bold face we indicate sources for which we performed X-ray data analysis using XMM-Newton observations} \\
\hline
\hline
\end{tabular}
\end{center}
\end{table*}

\section{Optical spectroscopy}

In this section, we describe the optical follow up work done
on all 29 objects. We list in Table \ref{log} the coordinates of 
the optical counterparts as obtained from the 2MASS catalogue\footnote{available 
at {\tt http://www.ipac.caltech.edu/2mass/}} (Skrutskie et al. 2006). In all but one case, these optical counterparts coincide in position 
with a galaxy, immediately suggesting that the majority of the objects listed
in Table \ref{log}
are active galaxies (their name as derived by NED\footnote{{\tt http://ned.ipac.caltech.edu/}} is reported in Tables \ref{agn1}, \ref{agn2}). The only exception is PBC J0826.3-7033, 
which is listed in NED and SIMBAD\footnote{{\tt http://simbad.u-strasbg.fr/simbad/ }} as an unidentified X-ray source; our optical follow up work locates this source at z=0 and thus 
demonstrates that it is not an extragalactic object but rather a galactic high energy source. 

Of the 28 AGN, 10 objects have an already known redshift, although not supported by published optical spectra, 9 have redshifts coming from SDSS\footnote{{\tt 
http://www.sdss.org}} (SDSS, Adelman-McCarthy et al. 2007) and 6dF (6dFGS; Jones et al. 
2004) archives and 9 have redshifts obtained from our spectroscopic observations are therefore published here for the first time.
In one case (a 6dF spectrum), our result is significantly different from  the one reported in the literature (see section 4) suggesting that it is important to confirm published 
redshift values especially for newly discovered objects.

Concerning the optical class for all our sample, 12 objects had an optical class
already reported in the Veron-Chetty \& Veron 
13th catalogue edition (V\&V13, Veron-Cetty \& Veron 2010 and references therein) and/or in NED.  
Nevertheless, we chose to report data on these 12 sources for a number of reasons: in a few cases our classification is  different or more detailed
than the one reported in the literature, and for a couple of sources different authors provide contradictory results and thus an ambiguous classification. 
For the remaining 17 objects, we reported the optical class supported by optical spectra for the first time.

\begin{table*}
\begin{center}
\caption[]{Main X-ray results concerning PBC J0826.3$-$7033.}\label{cvx}
\scriptsize
\resizebox{16cm}{!}{
\begin{tabular}{lccccc}
\noalign{\smallskip}
\hline
\hline
\noalign{\smallskip}
\multicolumn{1}{c}{Source} & \multicolumn{1}{c}{N$_{Hgal}$} & KT   &  $\chi^2/\nu$   & F$_{(2-10)keV}$ &F$_{(20-100)keV}$ \\
\noalign{\smallskip}
&$\times 10^{22}$cm$^{-2}$ & keV  & &$\times 10^{-11}$erg s$^{-1}$ cm$^{-2}$&$\times 10^{-11}$erg s$^{-1}$ cm$^{-2}$ \\
\noalign{\smallskip}
\hline
\noalign{\smallskip}
PBC J0826.3-7033    & 0.06$\pm$ 0.06&3.3$^{+1.9}_{-1.0}$  & 14.5/11 & 0.2 & 0.8 \\

\noalign{\smallskip}
\hline
\hline

\end{tabular}}
\end{center} 
\end{table*}

\begin{table*}[th!]
\caption[]{Main optical results concerning PBC J0826.3$-$7033
identified as a cataclismic variable.}
\label{cvo}
\hspace{-1.2cm}
\vspace{-.5cm}
\begin{center}
\resizebox{16cm}{!}{
\begin{tabular}{lcccccccccc}
\noalign{\smallskip}
\hline
\hline
\noalign{\smallskip}
\multicolumn{1}{c}{Object} & \multicolumn{2}{c}{H$_\alpha$} & 
\multicolumn{2}{c}{H$_\beta$} &
\multicolumn{2}{c}{He {\sc ii} $\lambda$4686} &
Optical & $A_V$ & $d$ & \multicolumn{1}{c}{$L_{\rm X}$} \\
\cline{2-7}
\noalign{\smallskip} 
 & EW & Flux & EW & Flux & EW & Flux & mag. & (mag) & (pc) & \\
\noalign{\smallskip}
\hline
\noalign{\smallskip}
PBC J0826.3$-$7033 & 38.9$\pm$1.8 & 66$\pm$3 & 33.6$\pm$1.5 & 44$\pm$2 & 5.7$\pm$0.9 & 7.4$\pm$1.1&
 13.8 ($R$) & 0 & 90 & 2 (2--10) \\
 & & & & & & & & & & 7(20--100) \\
\noalign{\smallskip}
\hline
\multicolumn{11}{l}{Note: EWs are expressed in \AA, line fluxes are
in units of 10$^{-15}$ erg cm$^{-2}$ s$^{-1}$, whereas X--ray luminosities
are in units of 10$^{30}$ erg s$^{-1}$} \\
\hline
\hline
\end{tabular}}
\end{center} 
\end{table*}

\begin{table*}[th!]
\begin{center}
\caption[]{Main results obtained from the analysis of the optical spectra of the 7 type 1 AGN}\label{agn1}
\scriptsize
\begin{tabular}{lccccccccc}
\noalign{\smallskip}
\hline
\hline
\noalign{\smallskip}
\multicolumn{1}{c}{Object} & $F_{\rm H_\alpha}$$^*$ & $F_{\rm H_\beta}$ &
$F_{\rm [OIII]}$ & Class & $z$ & \multicolumn{1}{c}{$D_L$} & \multicolumn{2}{c}{$E(B-V)$} & \multicolumn{1}{c}{NED Name}\\
\cline{8-9}
\noalign{\smallskip}
& & & & & & (Mpc) & Gal. & AGN& \\
\noalign{\smallskip}
\hline
\noalign{\smallskip}

PBC J0503.0+2300& 699$\pm$55 &113$\pm$23& 93$\pm$7& Sy1.5 & 0.058 & 259.3 & 0.515  &0.424 & 2MASX J05025822+2259520        \\
& [2670$\pm$153] &[600$\pm$64]  & [436$\pm$45] & & & & & &\\

& & & & & & & & & \\

PBC J0543.6$-2738$  & 83.9$\pm$18.4 & 107$\pm$17 & 22.6$\pm$3.1 & Sy1.2 & 0.009 & 38.8 & 0.029 & 0&ESO 424-G012                  \\
& [88$\pm$14.9]& [120$\pm$9] & [25.4$\pm$3.2] & & & & & &\\

& & & & & & & & & \\

PBC J0814.4+0421& 394$\pm$35 & 43.9$\pm$7.7 &31.9$\pm$2.3 & NLS1 & 0.034 & 149.4 & 0.027 & 1.073& CGCG 031-072                    \\
& [433$\pm$37] & [49.5$\pm$7.8] & [36.3$\pm$5.6] & & & & & & \\

& & & & & & & & & \\

PBC J1345.4+4141 &27.3$\pm$1.5 & 0.7$\pm$0.2& 3.7$\pm$1.8 & Sy1.9 & 0.009 & 37.1 & 0.007& 2.716& NGC 5290                        \\
& [33.9$\pm$1.9] & [0.7$\pm$0.2]& [3.7$\pm$1.8] & & & & & &\\

& & & & & & & & &\\

PBC J1439.0+1413 & 16.5$\pm$3.1 & --& --& Sy1.9 & 0.072& 325.1 & 0.019&--& 2MASX J14391186+1415215         \\
& [17.7$\pm$3.2]& &  & & & & & &\\

& & & & & & & & &\\

PBC J1453.0+2553& 470$\pm$68 & 111$\pm$22 & 19.7$\pm$3.3 & Sy1 & 0.049 & 217.7 & 0.039 & 0.406 &2MASX J14530794+2554327         \\
& [502$\pm$72] & [115$\pm$22] & [20.6$\pm$3.3] & & & & & &\\

& & & & & & & & &\\

PBC J1546.5+6931 & 181$\pm$20  & 26$\pm$5 & 97.8$\pm$8.6 & Sy1.9  & 0.037 & 162.9 & 0.041 & 1.036 & 2MASX J15462424+6929102        \\
& [244$\pm$23] & [29$\pm$5] & [113$\pm$16] & & & & & &\\

& & & & & & & & &\\

\noalign{\smallskip} 
\hline
\noalign{\smallskip} 
\multicolumn{10}{l}{Note: emission line fluxes are reported both as
observed and (between square brackets) corrected for the intervening Galactic} \\ 
\multicolumn{10}{l}{absorption $E(B-V)_{\rm Gal}$ along the object line of sight 
(from Schlegel et al. 1998). Line fluxes are in units of 10$^{-15}$ erg cm$^{-2}$ s$^{-1}$,} \\
\multicolumn{10}{l}{The typical error on the redshift measurement is $\pm$0.001 
except  for the SDSS and 6dFGS spectra, for which an uncertainty} \\
\multicolumn{10}{l}{of $\pm$0.0003 can be assumed.} \\
\multicolumn{10}{l}{$^*$: blended with [N {\sc ii}] lines} \\
\noalign{\smallskip} 
\hline
\hline
\end{tabular} 
\end{center}
\end{table*}

\begin{table*}[th!]
\begin{center}
\caption[]{Main results obtained from the analysis of the optical spectra of the 21 type 2 AGN }\label{agn2}
\scriptsize
\begin{tabular}{lccccccccc}
\noalign{\smallskip}
\hline
\hline
\noalign{\smallskip}
\multicolumn{1}{c}{Object} & $F_{\rm H_\alpha}$ & $F_{\rm H_\beta}$ &
$F_{\rm [OIII]}$ & Class & $z$ & \multicolumn{1}{c}{D$_L$} & \multicolumn{2}{c}{$E(B-V)$} & NED Name\\
\cline{8-9}
\noalign{\smallskip}
& & & & & & (Mpc) & Gal. & AGN & \\
\noalign{\smallskip}
\hline
\noalign{\smallskip}

PBC J0041.6+2534  & 11.3$\pm$4.5 & --& -- & Sy2/LINER & 0.015 & 65.0 & 0.035& --&NGC 214 \\
& [12.1$\pm$3.8] & & & & & & & & \\

& & & & & & & & &\\

PBC J0100.6$-$4752 & 37$\pm$4 & 10.8$\pm$2.9 & 101$\pm$6 & Sy2 & 0.048 & 213.1 & 0.013 &0.051& ESO 195-IG021           \\
& [38$\pm$4] & [12.6$\pm$3] & [106$\pm$6]  & & & & & &\\

& & & & & & & & &\\

PBC J0122.3+5004 & 139$\pm$23 & 34.1$\pm$5.7 & 169$\pm$28 & Sy2 & 0.021 & 91.4 & 0.217 &0.171& MCG +08-03-018                 \\
& [188$\pm$31]& [55$\pm$9] & [270$\pm$50] & & & & & &\\

& & & & & & & & & \\

PBC J0140.4$-$5320 & 27.3$\pm$4.3 & 4.6$\pm$0.9 &39.1$\pm$1.3 & Sy2 & 0.072 & 325.1 & 0.029 & 0.687 & 2MASX J01402676-5319389        \\
& [31.6$\pm$4.8] & [5.4$\pm$0.9] & [42.7$\pm$1.4] & & & & & &\\

& & & & & & & & &\\

PBC J0248.9+2627 & 47$\pm$4 & 4.2$\pm$0.5 & 21.6$\pm$1.5 & Sy2 & 0.057 & 274.3 & 0.158 & 1.221 &2MASX J02485937+2630391         \\
& [64.3$\pm$15.7] & [6.3$\pm$0.9] & [34.4$\pm$2.5] & & & & & &\\

& & & & & & & & &\\

PBC J0353.5+3713 & 35.5$\pm$2.8 & 6$\pm$1 & 12.8$\pm$1.3 & LINER & 0.019 & 82.6 & 0.536 & 0.105 & 2MASX J03534246+3714077        \\
& [120$\pm$13] & [37.6$\pm$6.4] & [71.4$\pm$13.9] & & & & & &\\

& & & & & & & & &\\

PBC J0356.9$-$4040  & 75.6$\pm$11.8& 21.7$\pm$5.6 & 167$\pm$9 & Sy2 & 0.075 & 65.0 & 0.035& 0.086&  2MASX J03565655-4041453        \\
& [70.1$\pm$7.2] &[22.4$\pm$5.6]  &[170$\pm$9]  & & & & & &\\

& & & & & & & & &\\

PBC J0544.3+5905& 3.9$\pm$0.5 & 0.7$\pm$0.1 &7.1$\pm$0.4 & Sy2 &  0.068 & 306.2 & 0.274&0.451 & 2MASX J05442257+5907361         \\
& [6.4$\pm$0.8] & [1.4$\pm$0.2] & [15.9$\pm$0.8] & & & & & &\\

& & & & & & & & &\\

PBC J0623.8$-$3212 & 97.9$\pm$17.4 &-- & 783$\pm$40 & Sy2& 0.035& 153.9 & 0.049 & --& ESO 426-G002                 \\
&[112$\pm$13.6 ]   & & [908$\pm$47] & & & & & &\\

& & & & & & & & &\\

PBC J0641.3+3251& 51.9$\pm$6.4 & 8.2$\pm$1.7 & 197$\pm$6.2 & Sy2 & 0.049 & 217.7 & 0.153 &0.577 &2MASX J06411806+3249313         \\
& [69.4$\pm$10.4] & [13.3$\pm$2.1] & [311$\pm$20] & & & & & &\\

& & & & & & & & &\\

PBC J0759.9+2324&  8.9$\pm$1.1& 1.1$\pm$0.3 & 10.6$\pm$1.1 & Sy2  & 0.029 & 127 & 0.059 & 0.859 & CGCG 118-036                   \\
& [9.8$\pm$1.1]& [1.4$\pm$0.4] & [13.5$\pm$1.2] & & & & & &\\

& & & & & & & & &\\

PBC J0919.9+3712  &4.3$\pm$0.4  &0.5$\pm$0.1 &4.1$\pm$0.3 & Sy2 & 0.0075 & 32.3 & 0.012&1.079 & IC 2461                         \\
&[4.4$\pm$0.4] & [0.5$\pm$0.1] &4.2$\pm$0.3 & & & & & &\\

& & & & & & & & &\\

PBC J0954.8+3724& 8.6$\pm$0.5& 0.8$\pm$0.03 &3.6$\pm$0.6 & Sy2  & 0.019 & 82.6 & 0.016 & 1.292& IC 2515                         \\
&[8.8$\pm$0.6]& [0.8$\pm$0.2] & [3.7$\pm$0.5] & & & & & &\\

& & & & & & & & &\\

PBC J1246.9+5432& -- & --& 16.4$\pm$4.2& Sy2 & 0.017 & 73.8 & 0.017 & --& NGC 4686                        \\
&  & & [17.8$\pm$4.4] & & & & & &\\

& & & & & & & & &\\

PBC J1335.8+0301 &19.6$\pm$2.4 & 2.9$\pm$0.5 & 16.9$\pm$1.3 & Sy2  & 0.0218 & 94.9 & 0.024 &1.014& NGC 5231                        \\
& [20.7$\pm$1.9]& [3.2$\pm$0.5] & [17.9$\pm$1.1] & & & & & &\\

& & & & & & & & &\\

PBC J1344.2+1934 &16.6$\pm$1.8 & --& 6.6$\pm$1.1 & Sy2/LINER & 0.027 & 118 & 0.027& -- & CGCG 102-048                    \\
& [17.2$\pm$1.8] & & [6.9$\pm$1.2] & & & & & &\\

& & & & & & & & &\\

PBC J1506.6+0349 & 17.2$\pm$1.1& 2.5$\pm$0.6 & 20.8$\pm$0.8 & Sy2  & 0.038 & 167.5 & 0.049 &0.884 & 2MASX J15064412+0351444        \\
& [19.4$\pm$2.6]& [2.7$\pm$0.7]& [$23.7\pm$1.4] & & & & & &\\
& & & & & & & & &\\

PBC J1620.3+8101 & 2.4$\pm$0.8 & 2.5$\pm$0.6 & 19.2$\pm$3.9 & Sy2  & 0.025 & 109.1& 0.046 &0.832 &  CGCG 367-009        \\
& [27.9$\pm$7.8]& [4.1$\pm$1.5]& [23.9$\pm$4.5] & & & & & &\\

& & & & & & & & &\\

PBC J2148.2$-$3455 & 6460$\pm$582& 857$\pm$83  &4970$\pm$347& Sy2  & 0.0161 & 70.7 & 0.029 & 0.832 &NGC 7130                         \\
& [6900$\pm$895]& [947$\pm$117]&[5440$\pm$347]& & & & & &\\

& & & & & & & & &\\

PBC J2333.9$-$2343 & -- & 3.2$\pm$1.2 &14.7$\pm$2.8 & Sy2& 0.0475 & 210.8 & 0.029& -- & PKS 2331-240                    \\
& & [3.5$\pm$1.2] &[16.3$\pm$2.8 ] & & & & & &\\

& & & & & & & & &\\

PBC J2341.9+3036 & 9.7$\pm$4.6 & -- & 8.3$\pm$2.1& Sy2  & 0.017 & 73.8 & 0.102 &-- & UGC 12741                      \\
&  [14.2$\pm$5.3]& & [15.6$\pm$2.9] & & & & & &\\

& & & & & & & & &\\

\noalign{\smallskip} 
\hline
\noalign{\smallskip} 
\multicolumn{10}{l}{Note: emission line fluxes are reported both as 
observed and (between square brackets) corrected for the intervening Galactic} \\ 
\multicolumn{10}{l}{absorption $E(B-V)_{\rm Gal}$ along the object line of sight 
(from Schlegel et al. 1998). Line fluxes are in units of 10$^{-15}$ erg cm$^{-2}$ s$^{-1}$,} \\
\multicolumn{10}{l}{The typical error on the redshift measurement is $\pm$0.001 
except for the SDSS and 6dFGS spectra, for which an uncertainty} \\
\multicolumn{10}{l}{of $\pm$0.0003 can be assumed.} \\
\noalign{\smallskip} 
\hline
\hline
\end{tabular} 
\end{center}
\end{table*}

In Table~\ref{log} the detailed log of all optical  measurements is also reported: we list in column 7  the telescope and  instrument 
used for the observation while the characteristics of each spectrograph are given in columns 8 and 9. Column 10 provides the 
observation date and the UT time at mid-exposure, while column 11 reports 
the exposure times and the number of spectral pointings.

For 20 sources the following telescopes were used for the optical spectroscopic study presented here:

\begin{itemize}
\item the 1.52m ``Cassini'' telescope of the Astronomical Observatory of 
Bologna, in Loiano, Italy;
\item the 1.9m ``Radcliffe'' telescope of the South African Astronomical 
Observatory (SAAO) in Sutherland, South Africa;
\item the 2.1m telescope of the Observatorio Astr\'onomico Nacional in San Pedro Martir, Mexico;
\item the 3.58m New Technology Telescope (NTT) at the ESO-La Silla Observatory, Chile; 
\end{itemize}
The data reduction was performed with the standard procedure (optimal extraction; Horne 1986) using IRAF\footnote{
IRAF is the Image Reduction and Analysis 
Facility made available to the astronomical community by the National 
Optical Astronomy Observatories, which are operated by AURA, Inc., under 
contract with the U.S. National Science Foundation. It is available at 
{\tt http://iraf.noao.edu/}}.
Calibration frames (flat fields and bias) were taken on the day preceding or following 
the observing night. The wavelength calibration was obtained using lamp spectra 
acquired soon after each on-target spectroscopic acquisition. The uncertainty on the 
calibration was $\sim$0.5~\AA~for all cases; this was checked using the positions of 
background night sky lines. Flux calibration was performed using 
catalogued spectrophotometric standards.
Objects with more than one observation had their spectra stacked 
together to increase the signal-to-noise ratio.

Further spectra (that is, 9 out of 29) were retrieved from two different astronomical archives:
the Sloan Digitized Sky Survey and 
the Six-degree Field Galaxy Survey\footnote{{\ 
http://www.aao.gov.au/local/www/6df/}}. As the 6dFGS archive provides spectra which are not calibrated in 
flux, we used the optical photometric information in Jones et al. (2005) 
and Doyle et al. (2005) to calibrate the 6dFGS data presented here.

The identification and classification approach we adopt in the analysis of the optical spectra is the following:
for the emission-line AGN classification, we used the criteria 
of Veilleux \& Osterbrock (1987) and the line ratio diagnostics 
of Ho et al. (1993, 1997) and of Kauffmann et al. (2003) to disentangle among Seyfert 2, starburst galaxies, H{\sc ii} regions and low-ionization nuclear emission-line regions (LINERs; Heckman 1980). In this last class, some lines ([O{\sc ii}]$_{\lambda 3723}$, [O{\sc i}]$_{\lambda 6300}$,
[N{\sc ii}]$_{\lambda 6584}$) are stronger than in typical Seyfert 2 galaxies; the permitted emission line luminosities are weak; and
the emission line widths are comparable with those of type 2 AGN.
In particular, as mentioned in the work of Ho et al. (1993), all  sources
 with [O{\sc ii}] $>$ [O{\sc iii}], [N{\sc ii}]/H$_{\alpha}$ $>$ 0.6, [O{\sc i}] $>$1/3 [O{\sc iii}] can be
considered LINERs (see Table \ref{agn2}). 
For the subclass assignation to Seyfert 1 galaxies, we used the 
\mbox{H$_\beta$/[O {\sc iii}]$\lambda$5007} line flux ratio criterion as in
Winkler et al. (1992). Moreover, the criteria of Osterbrock \& Pogge (1985) allowed us to discriminate between `normal' Seyfert 1 and narrow line Seyfert 1 (NLS1);
the latter are galaxies with a full width at half-maximum (FWHM) of the H$_{\beta}$ line lower than 2000 km s$^{-1}$, with 
permitted lines which are only slightly broader than the forbidden lines, with a [O{\sc iii}]$_{\lambda 5007}$/H$_{\beta}$ ratio $<$ 3 and finally with evident
Fe{\sc ii} and other high-ionization emission-line complexes.

Note that the spectra of all extragalactic objects are not corrected for starlight contamination 
(see, e.g., Ho et al. 1993, 1997), because of their limited S/N and  
spectral resolution. This however does not affect our results and conclusions.

To estimate the E(B-V) local optical absorption in our AGN sample, when possible,
we first dereddened the  H$_\alpha$ and H$_\beta$ line 
fluxes by applying a correction for the galactic absorption along the line of sight to the source. This was done 
using the galactic color excess $E(B-V)_{\rm Gal}$ given by Schlegel et al. (1998) and
the galactic extinction law obtained by Cardelli et al. (1989).
We then estimated the color excess  \mbox{$E(B-V)_{\rm AGN}$} local to the AGN host galaxy by comparing the intrinsic 
line ratio and that corrected for galactic reddening using the following relation for type 2 AGNs as derived from Osterbrock (1989): 
$$E(B-V)= a\; Log \left(\frac{H_{\alpha}/H_{\beta}}{(H_{\alpha}/H_{\beta})_0}\right)$$ 
For type 1 objects, where the H$_{\alpha}$ is strongly blended with the forbidden narrow [N{\sc ii}] lines, it is not easy to obtain 
a reliable H$_{\alpha}$/H$_{\beta}$
estimate. In these cases, we have used the H$_{\gamma}$/H$_{\beta}$ ratio, taking into account that H$_{\gamma}$ may also be blended with 
the [O {\sc iii}]$_{\lambda 4363}$ line.
In the above relation H$_{\alpha}/H_{\beta}$ is the observed Balmer decrement, $(H_{\alpha}/H_{\beta})_0$ is the intrinsic one (2.86), while
{\it a} is a constant with a value of 2.21. When instead we use the H$_{\gamma}$, we adopted the same relation described before, using an intrinsic $(H_{\gamma}$/H$_{\beta})_0$ of 0.474 and with an {\it a} value of -5.17.

Finally, we estimated the mass of the central black hole for a few of the type 1 AGN found in the sample\footnote{We could not estimate the mass of the central black hole of PBC J1439.0+1413 because it lacks the H$_{\beta}$ emission line, and
for PBC J1345.4+4141 and PBC J1546.5+6931 because only the narrow component of the H$_{\beta}$ line was observed in their spectra.}.
The method used here follows the prescription of Wu et al. (2004) and Kaspi et al. (2000), 
where we use the H$_{\beta}$ emission line flux, corrected for the galactic color excess (Schlegel et al. 1998), and the broad-line region
(BLR) gas velocity ($v_{FWHM}$).  Through Eq. (2) of Wu et al. (2004) we estimate the BLR size, which is used with  $v_{FWHM}$ in Eq. (5) of Kaspi et al. (2000)
to calculate the AGN black hole mass. The results are reported in Table \ref{blr}. 

To derive the distance of the only compact galactic X-ray source of our sample, we use the distance modulus assuming
an absolute magnitude M$_{V} \sim$ +9 and an intrinsic color index (V-R)$_{0} \sim$ 0 mag (Warner 1995).
Although this method basically provides an order-of-magnitude value for the distance of this Galactic source, our past experience (Masetti et al. 2004, 2006a, 2006b, 2008, 2009, 2010) tells us that this estimate is in general correct to within 50\% of the refined value subsequently determined with more precise approaches.
To calculate instead the luminosity distances of the 28 galaxies in the sample, we consider a \mbox{cosmology} with $H_{\rm 0}$ = 70
km s$^{-1}$ Mpc$^{-1}$, $\Omega_{\Lambda}$ = 0.7 and $\Omega_{\rm m}$ =
0.3 and used the Cosmology
Calculator of Wright (2006).

\begin{figure}[th!]
\hspace{-.1cm}
\psfig{file=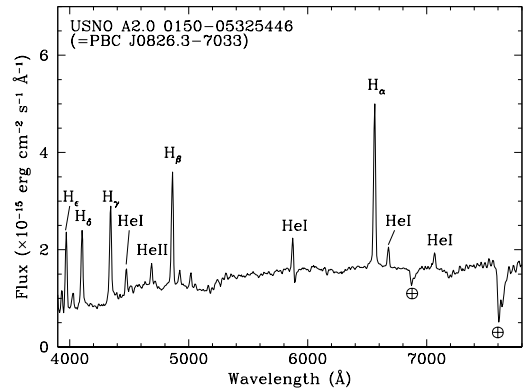,width=8cm,angle=270}
\caption{Spectrum (not corrected for the intervening galactic absorption) of the 
optical counterpart of the Galactic CV PBC J0826.3-7033.
}\label{cv}
\end{figure}

\begin{figure*}[th!]
\hspace{-.1cm}
\centering{\mbox{\psfig{file=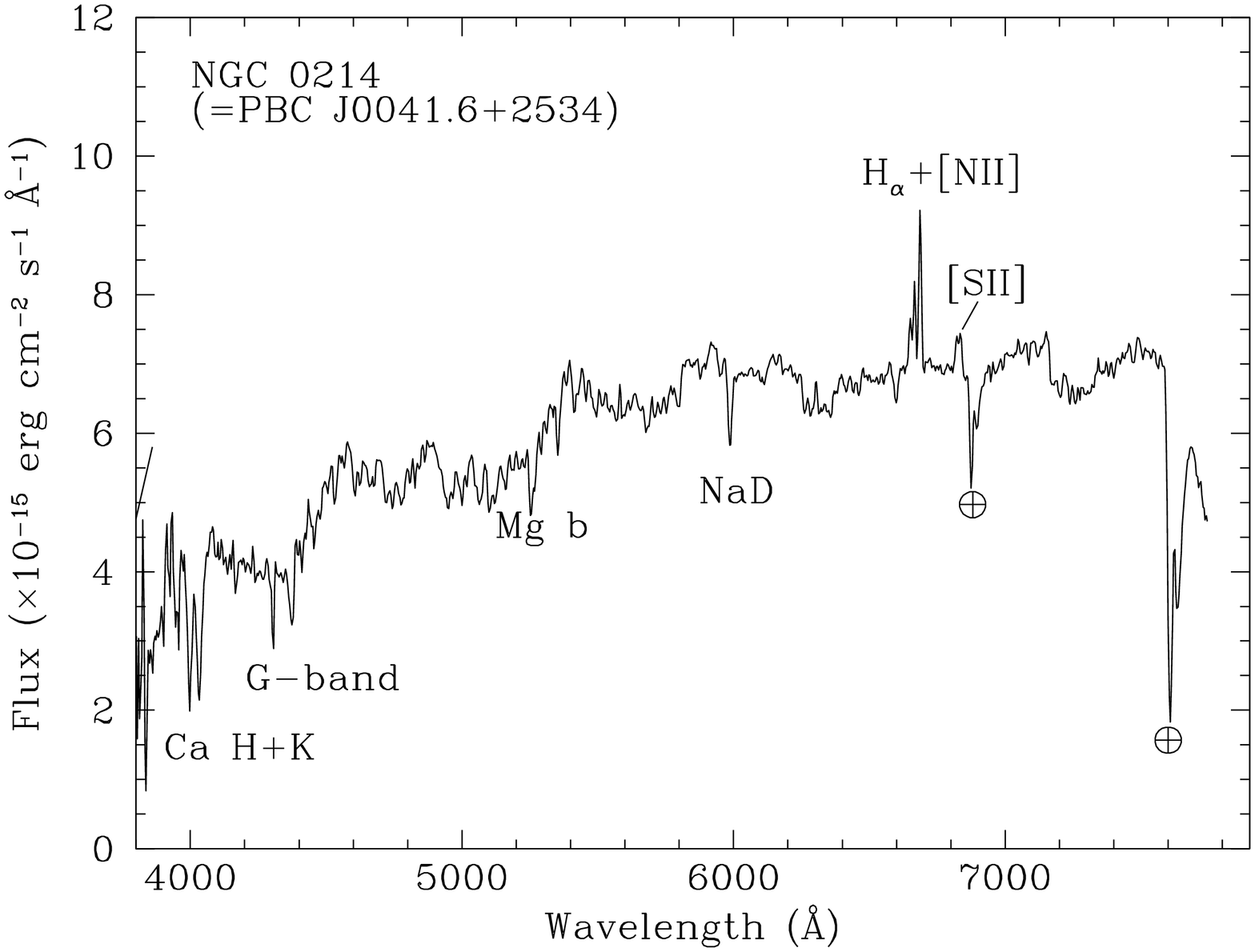,width=6.0cm,angle=270}}}
\centering{\mbox{\psfig{file=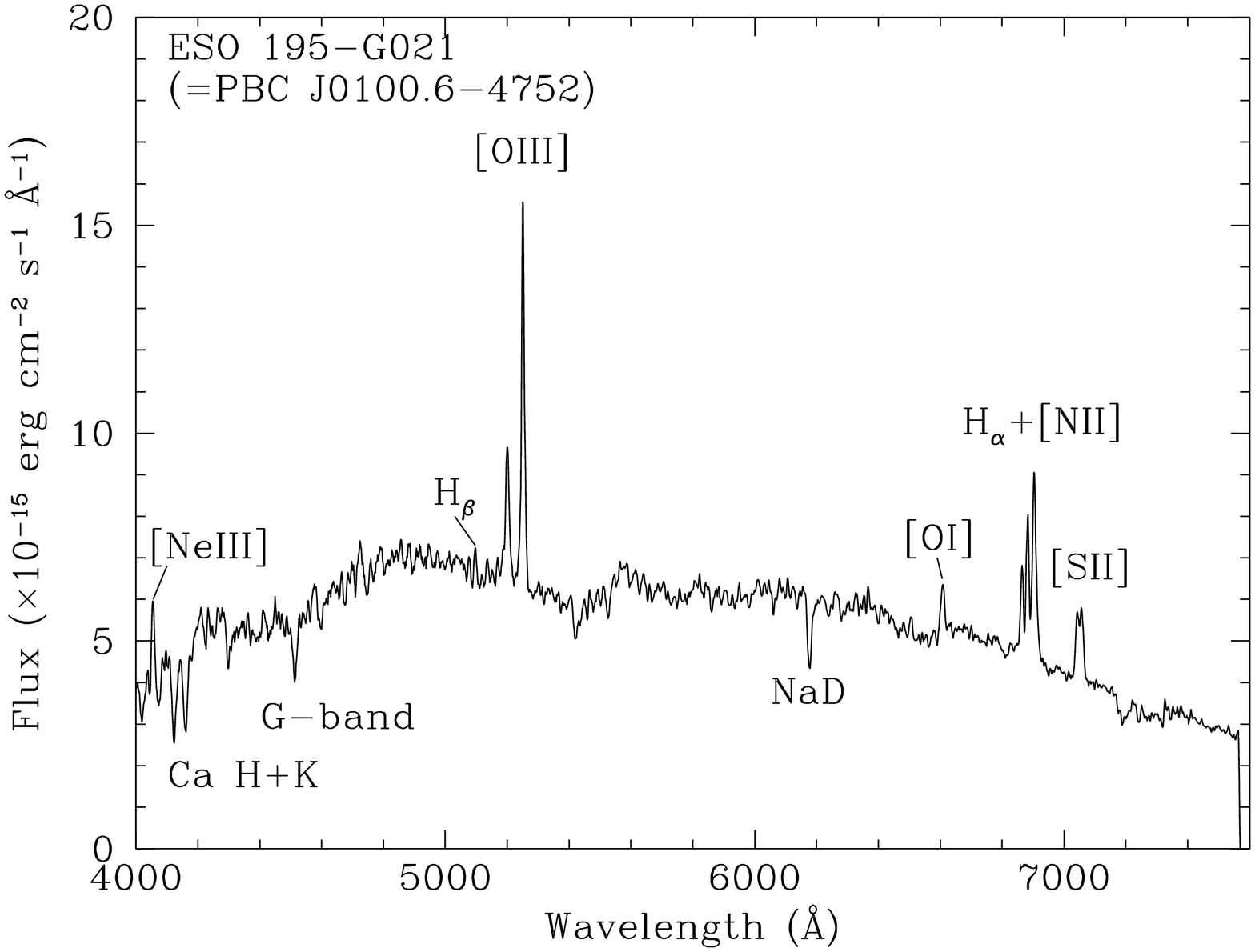,width=6.0cm,angle=270}}}
\centering{\mbox{\psfig{file=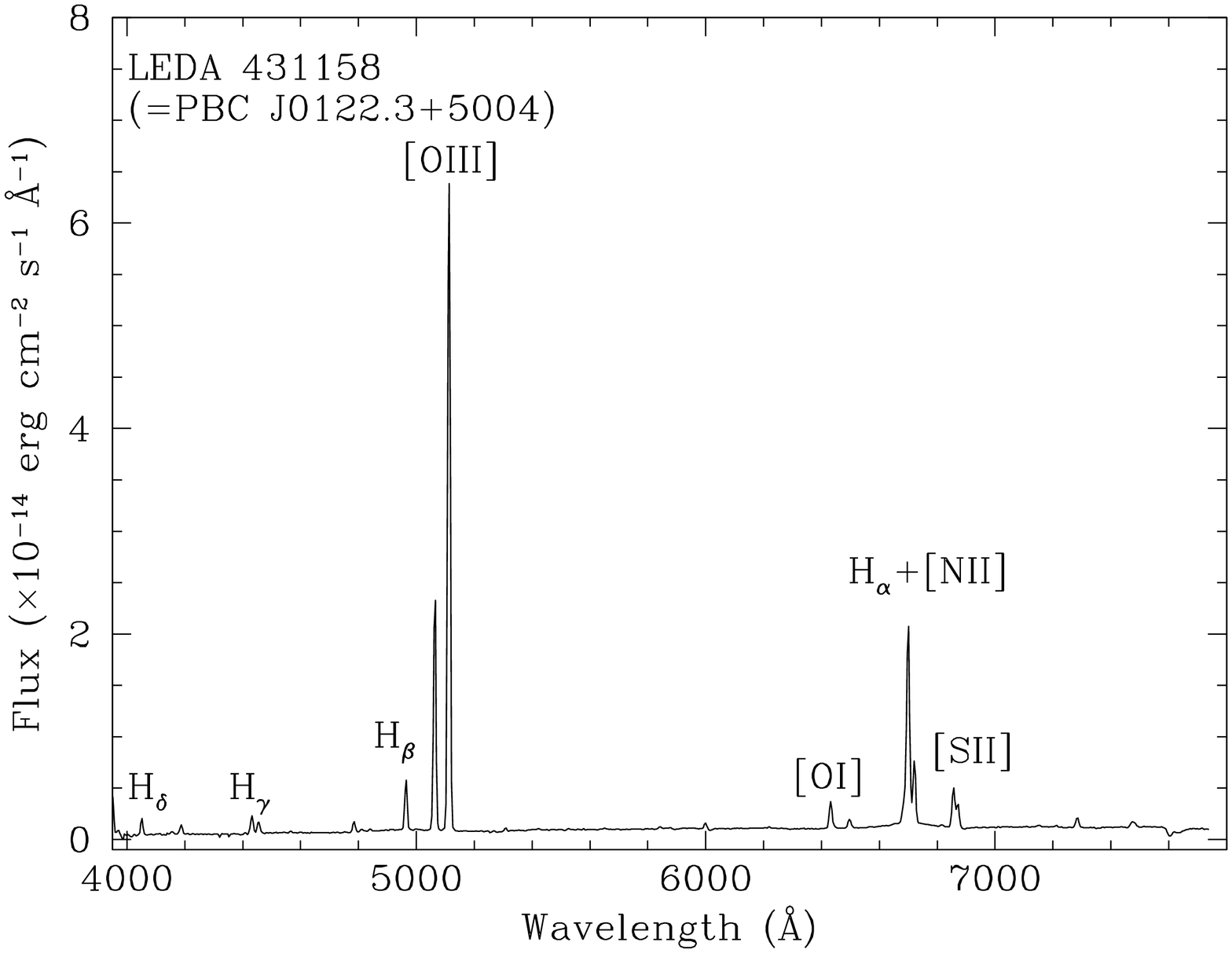,width=6.0cm,angle=270}}}
\centering{\mbox{\psfig{file=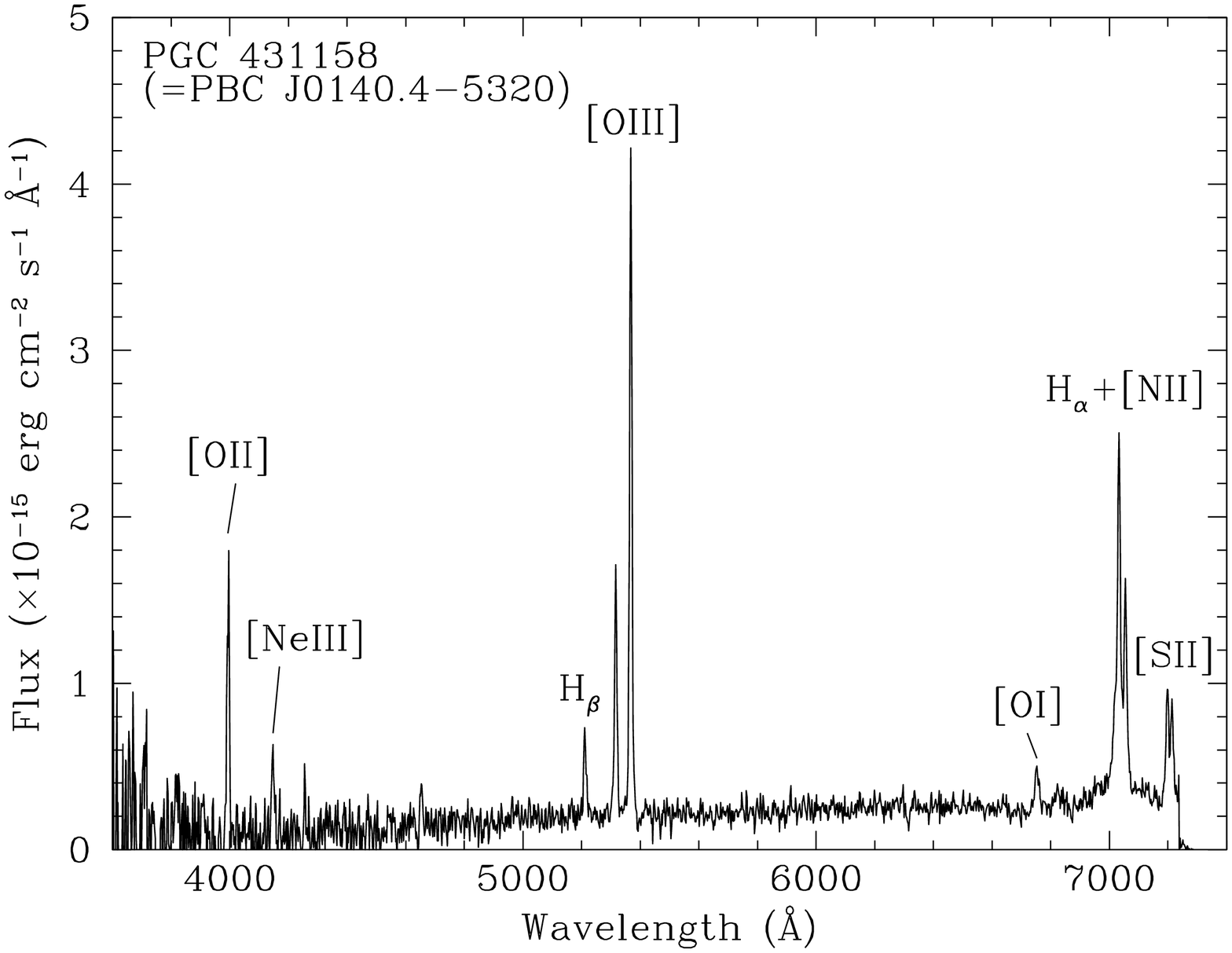,width=6.0cm,angle=270}}}
\centering{\mbox{\psfig{file=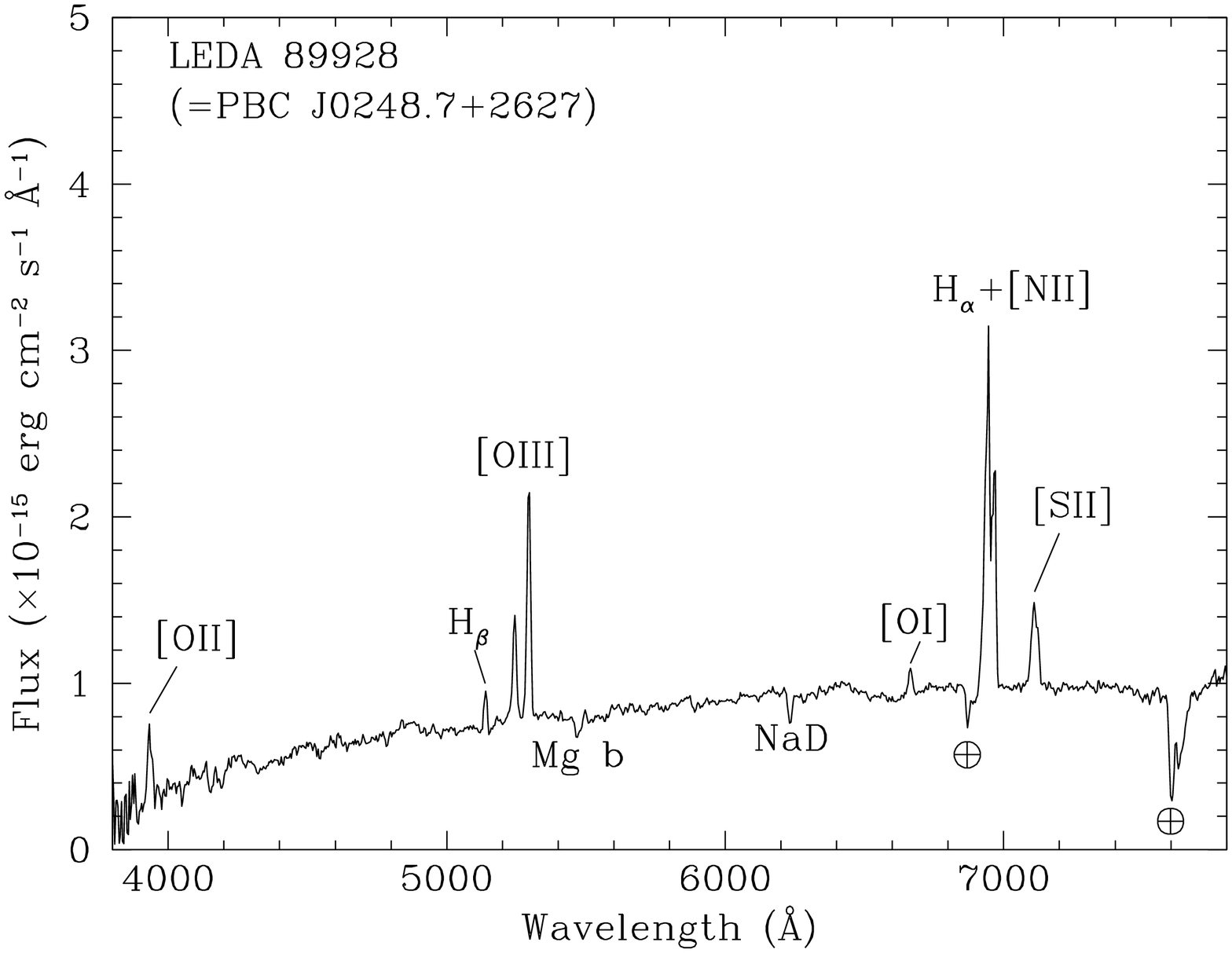,width=6.0cm,angle=270}}}
\centering{\mbox{\psfig{file=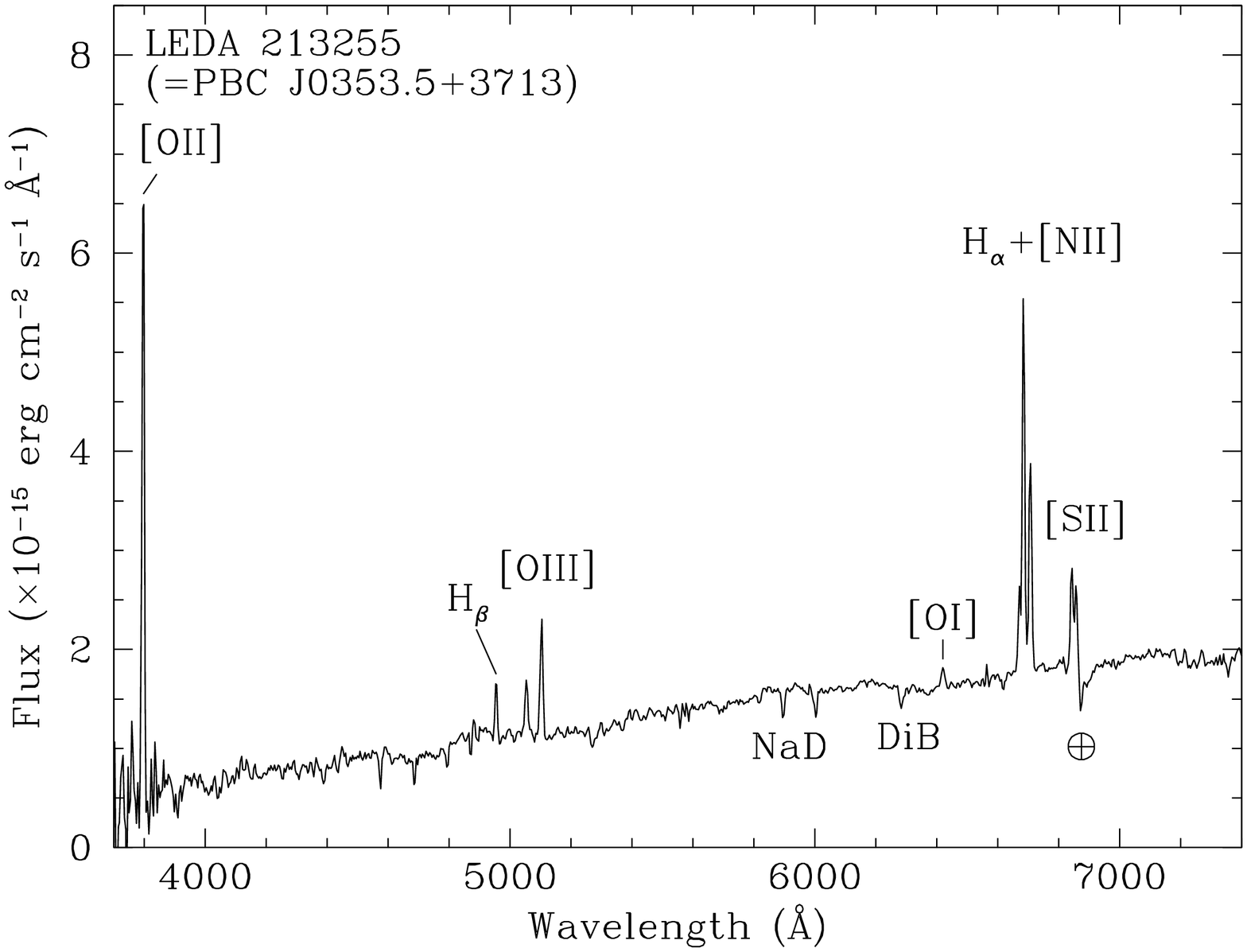,width=6.0cm,angle=270}}}
\centering{\mbox{\psfig{file=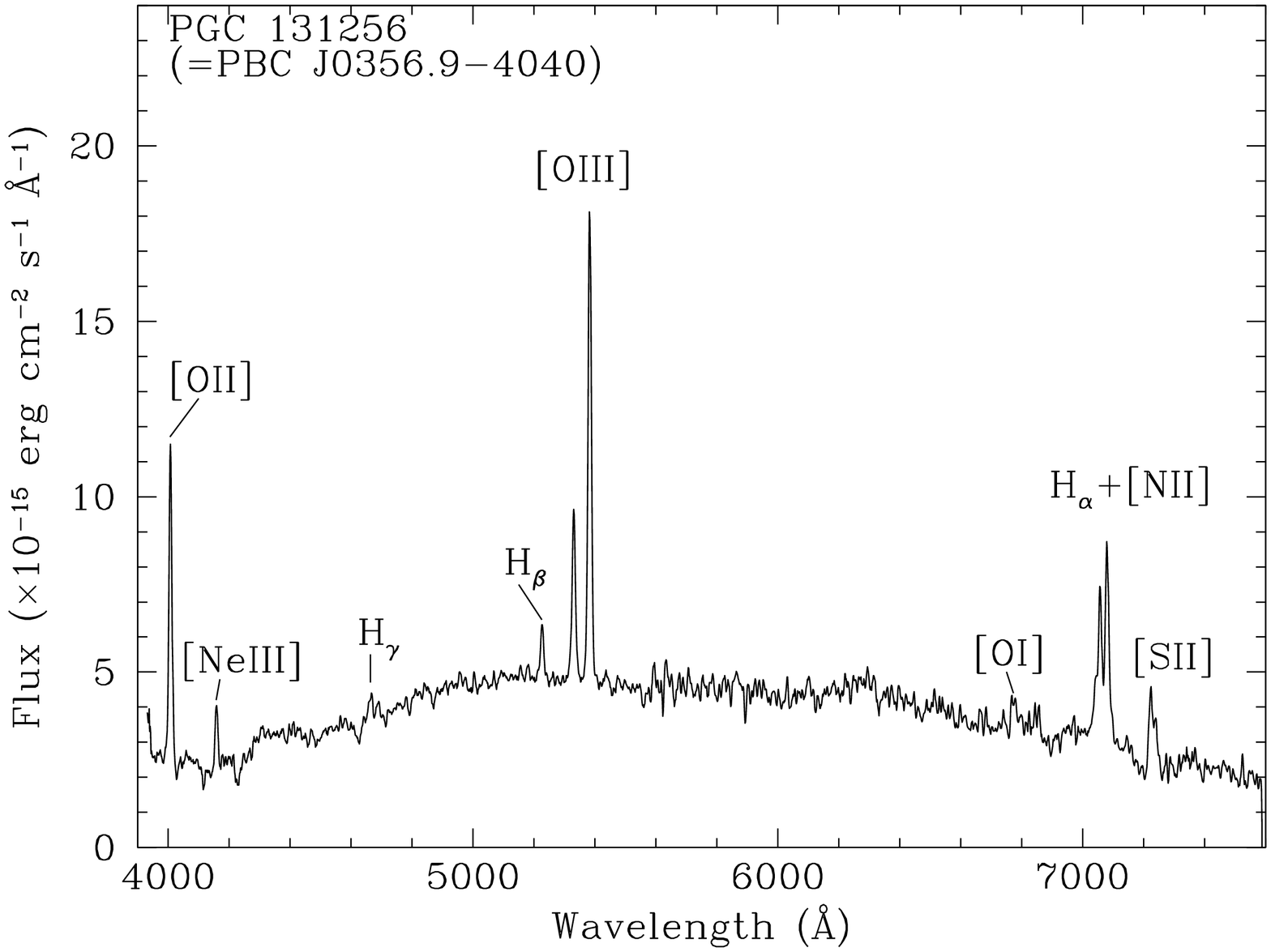,width=6.0cm,angle=270}}}
\centering{\mbox{\psfig{file=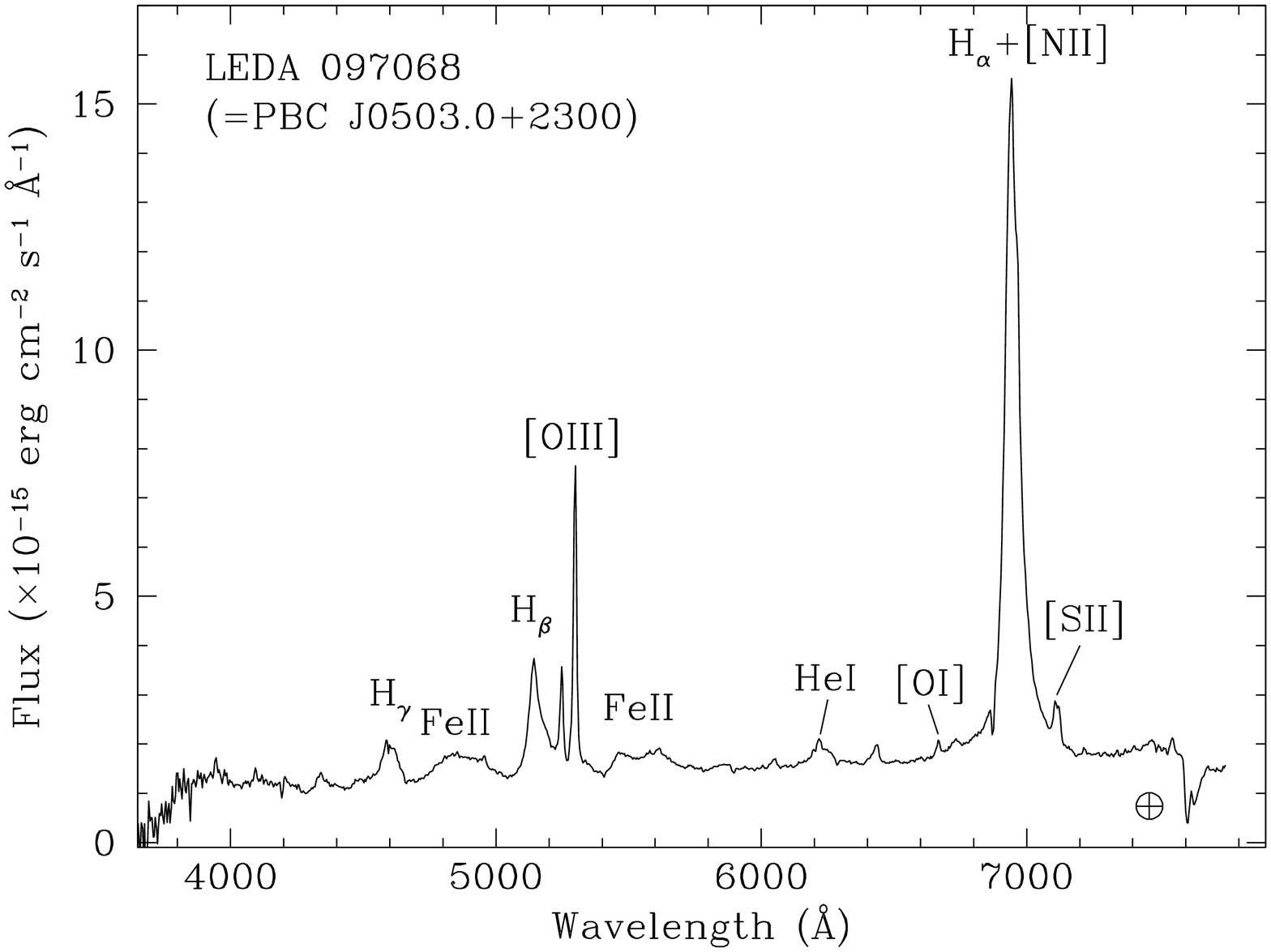,width=6.0cm,angle=270}}}
\centering{\mbox{\psfig{file=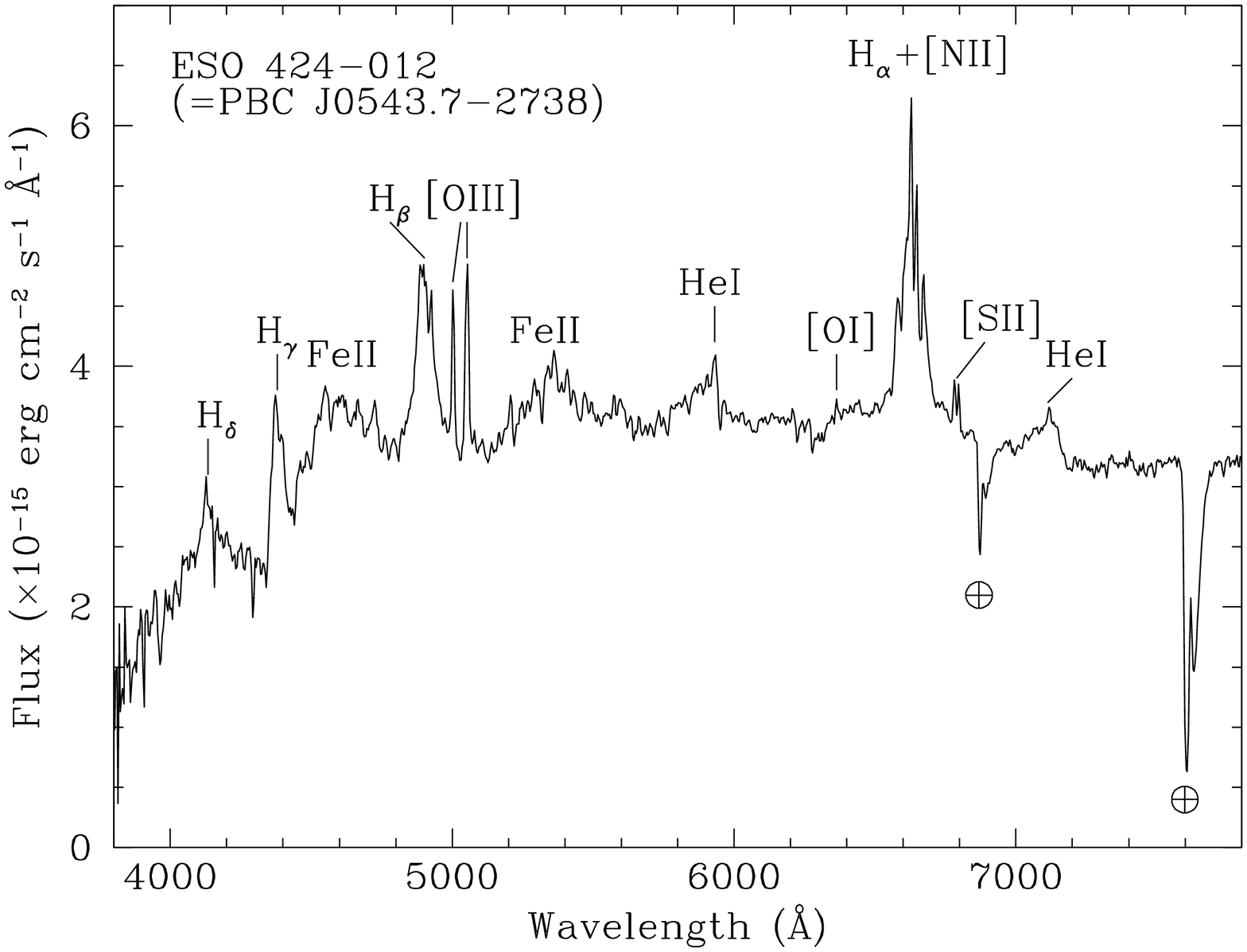,width=6.0cm,angle=270}}}
\centering{\mbox{\psfig{file=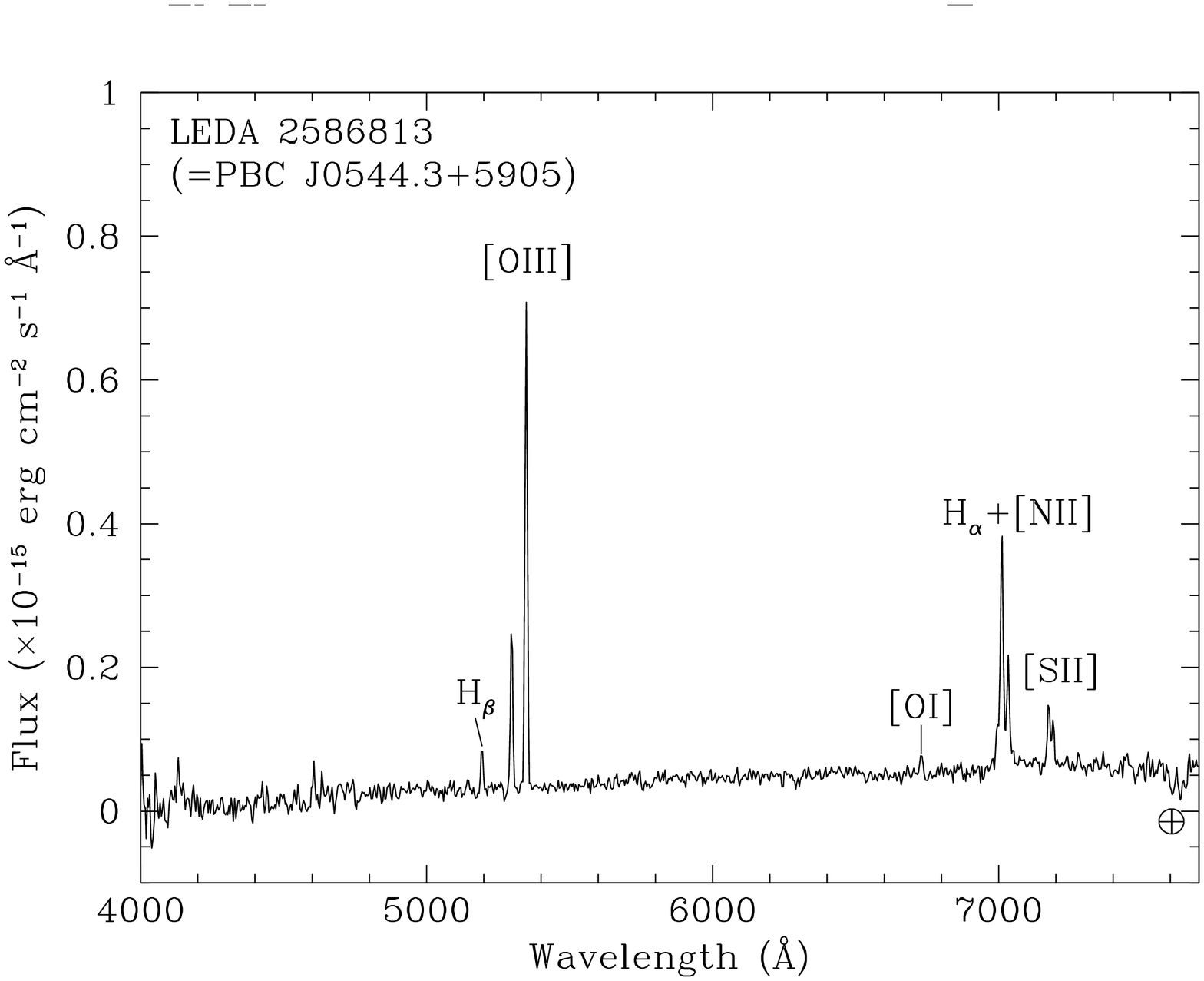,width=6.0cm,angle=270}}}
\centering{\mbox{\psfig{file=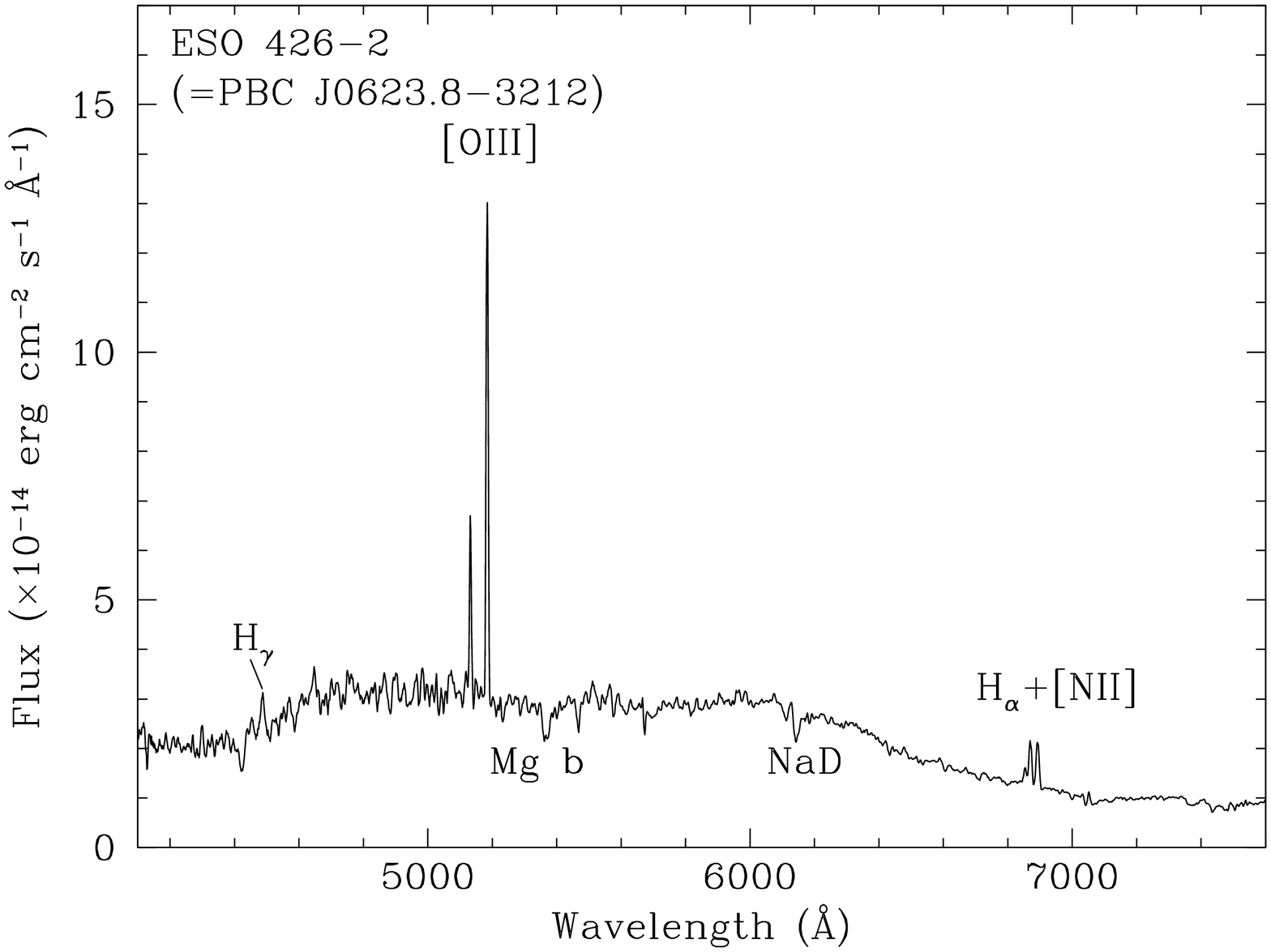,width=6.0cm,angle=270}}}
\centering{\mbox{\psfig{file=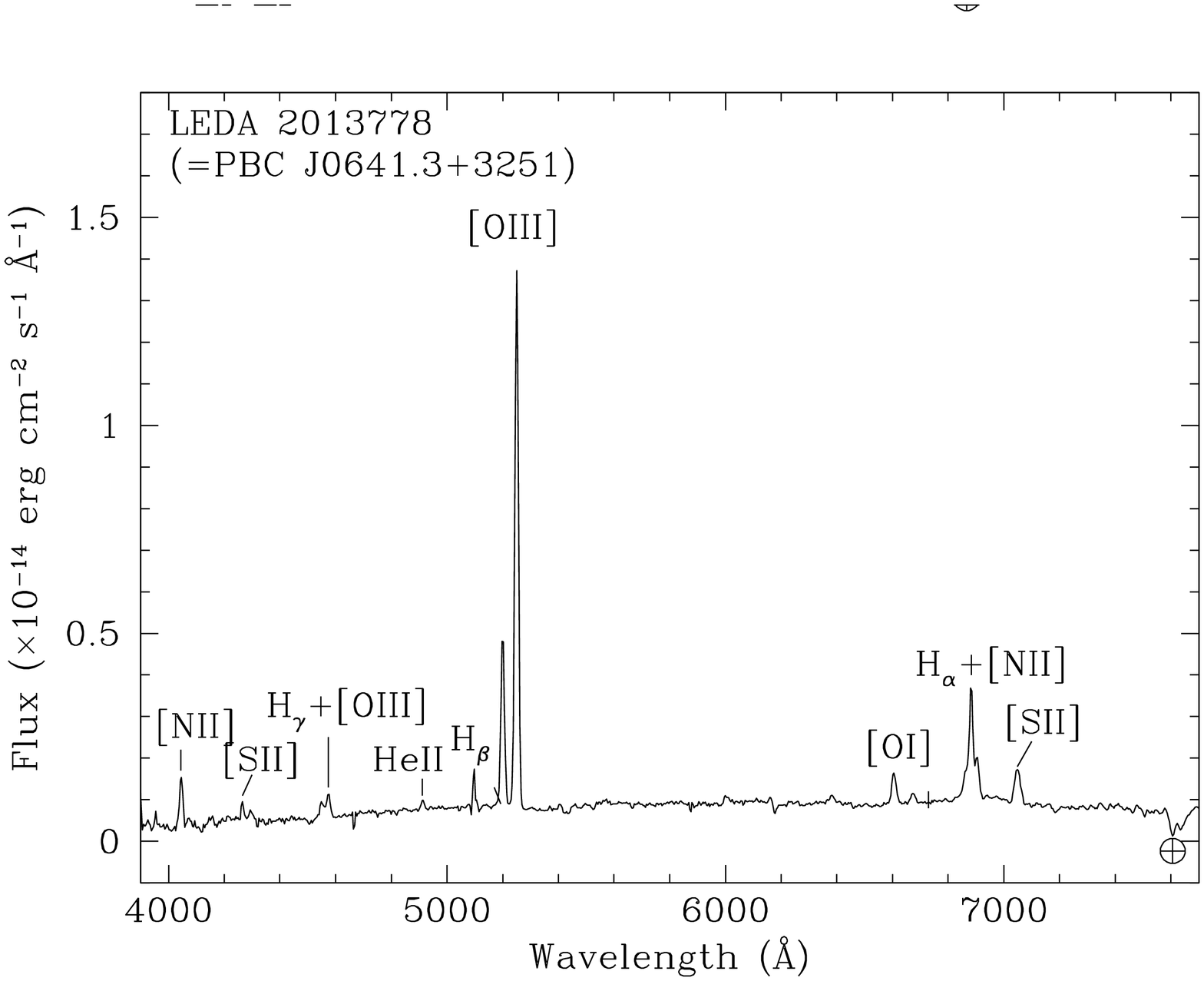,width=6.0cm,angle=270}}}
\centering{\mbox{\psfig{file=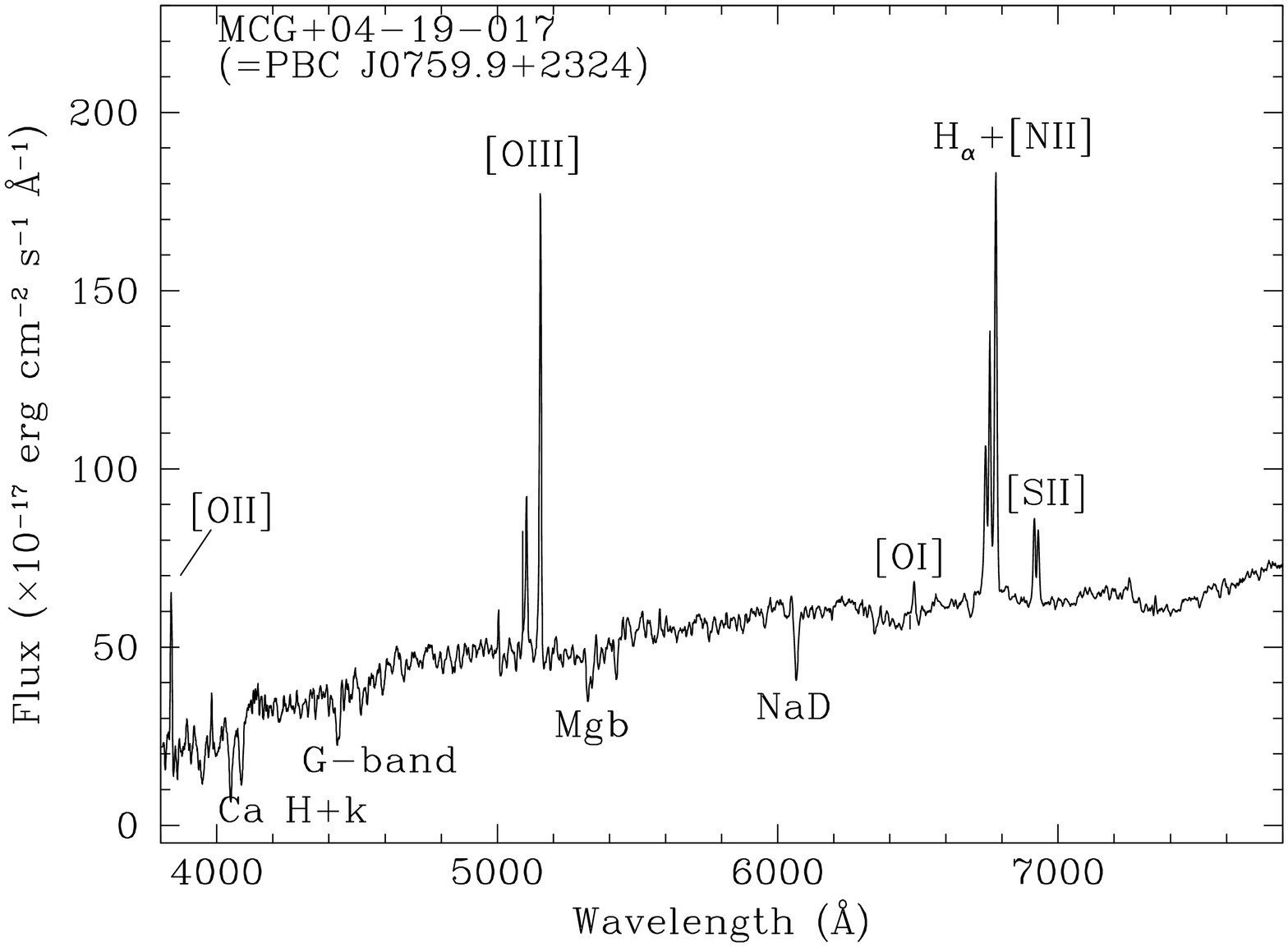,width=6.0cm,angle=270}}}
\centering{\mbox{\psfig{file=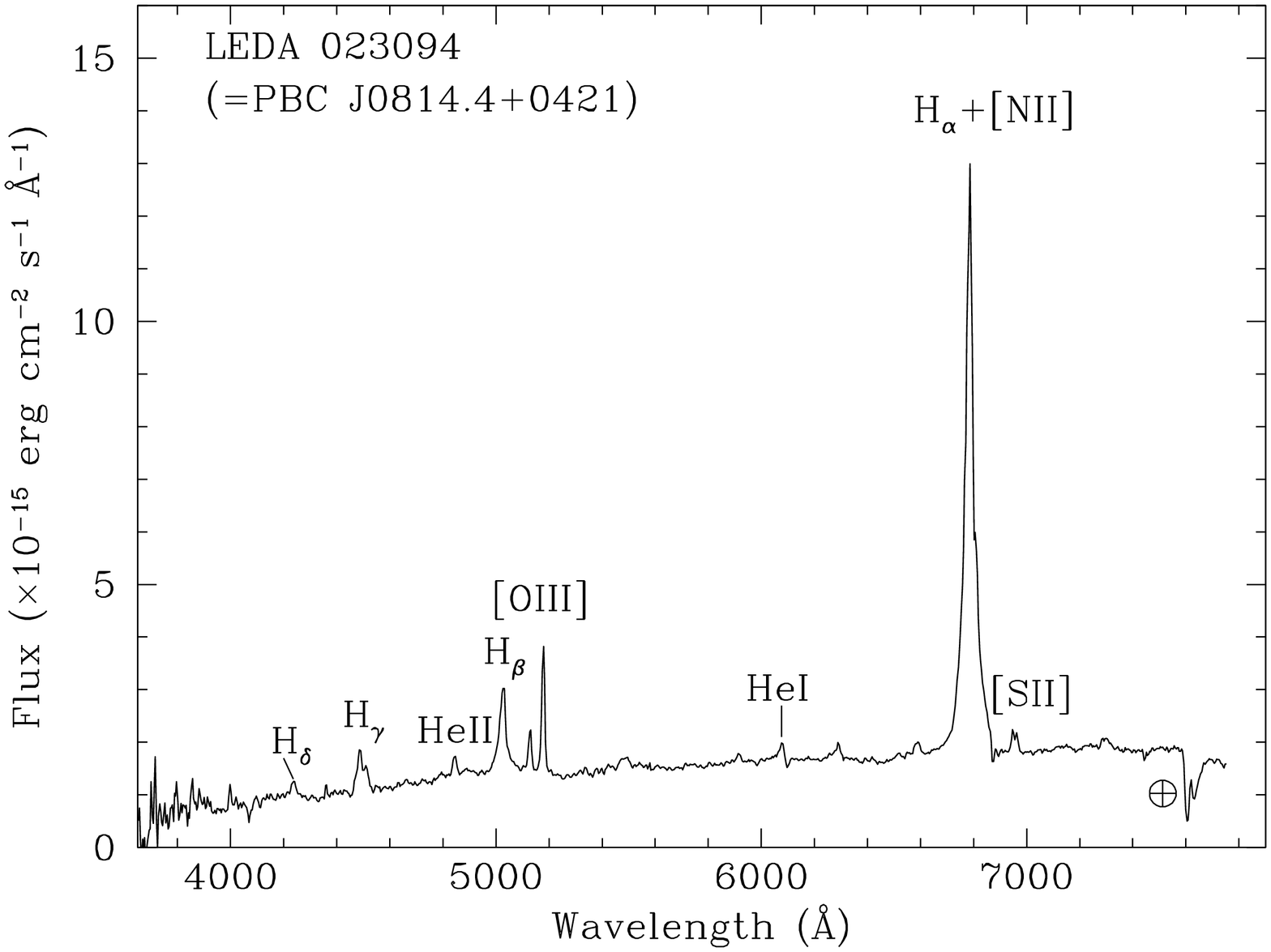,width=6.0cm,angle=270}}}
\centering{\mbox{\psfig{file=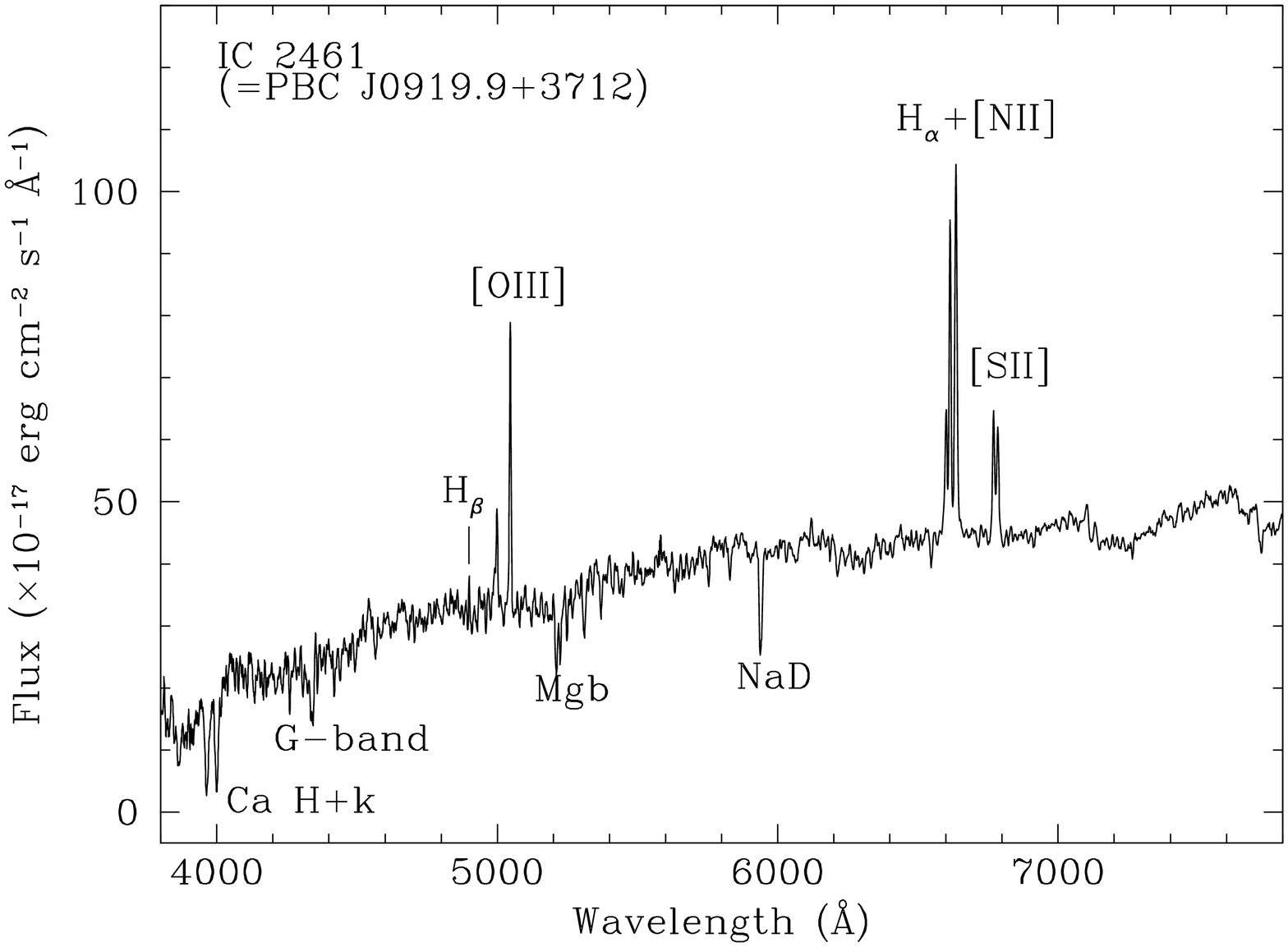,width=6.0cm,angle=270}}}
\caption{Spectra (not corrected for the intervening galactic absorption) of the 
optical counterpart of  PBC J0041.6+2534, PBC J0100.6$-$4752, PBC J0122.3+5004, PBC J0140.4$-$5320, PBC J0248.9+2627, PBC J0353.5+3713, PBC J0356.9$-$4040, PBC J0503.0+2300, PBC J0543.6$-$2738, PBC J0544.3+5905, PBC J0623.8$-$3212 and PBC J0641.3+3251, PBC J0759.9+2324, PBC J0814.4+0421 and PBC J0919.9+3712.
}\label{spectra1}
\end{figure*}

\begin{figure*}
\hspace{-.1cm}
\centering{\mbox{\psfig{file=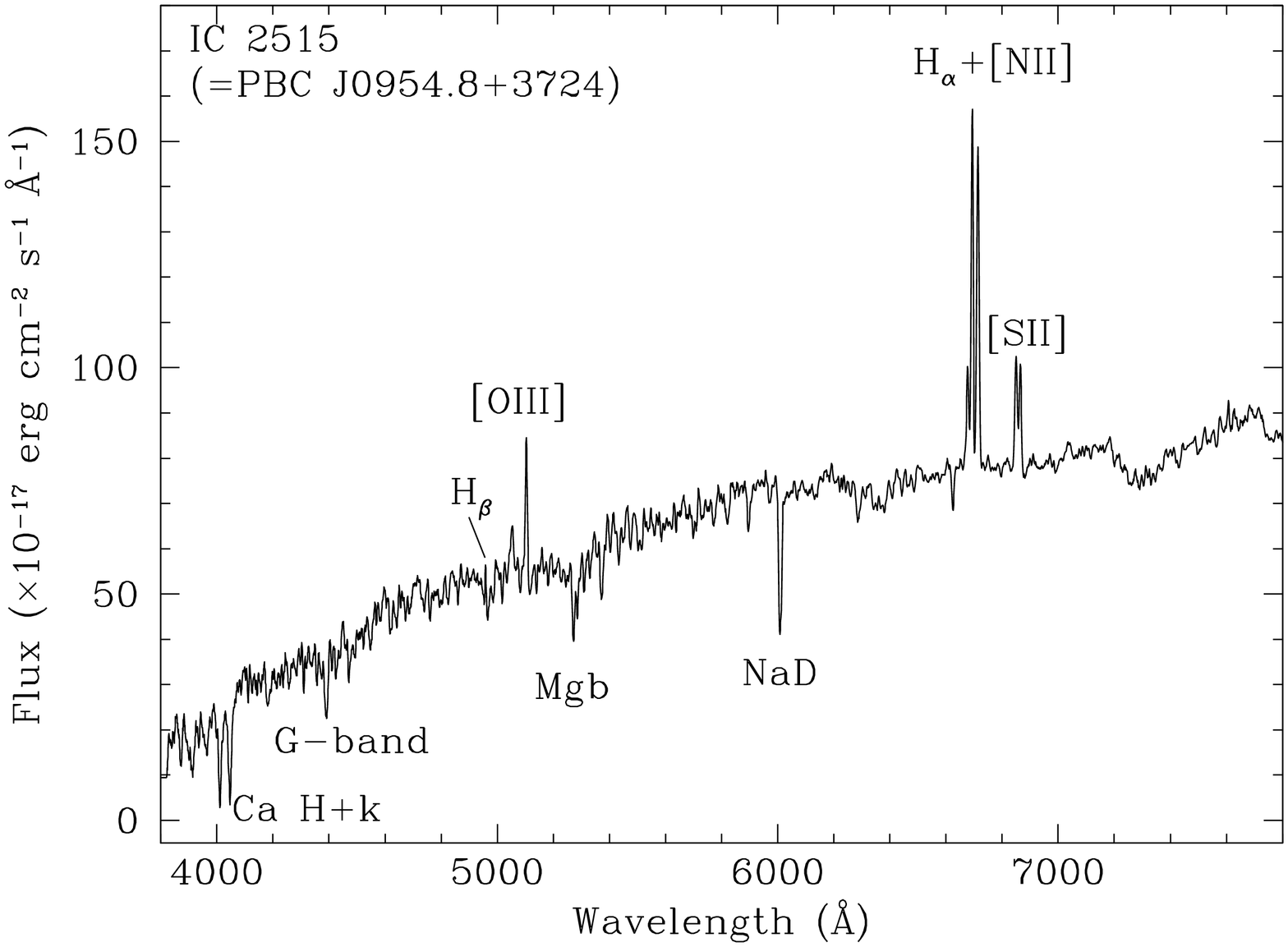,width=6.0cm,angle=270}}}
\centering{\mbox{\psfig{file=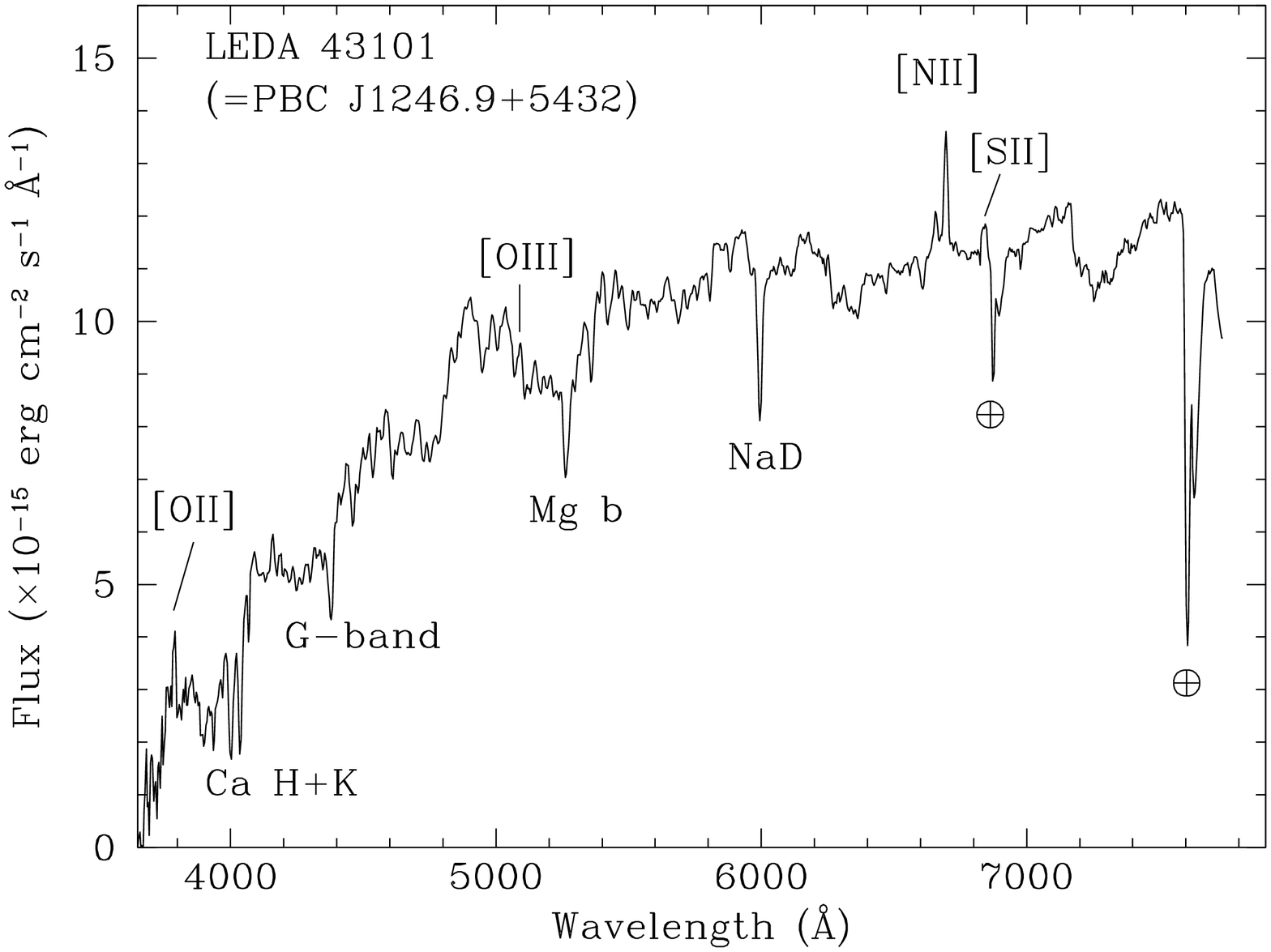,width=6.0cm,angle=270}}}
\centering{\mbox{\psfig{file=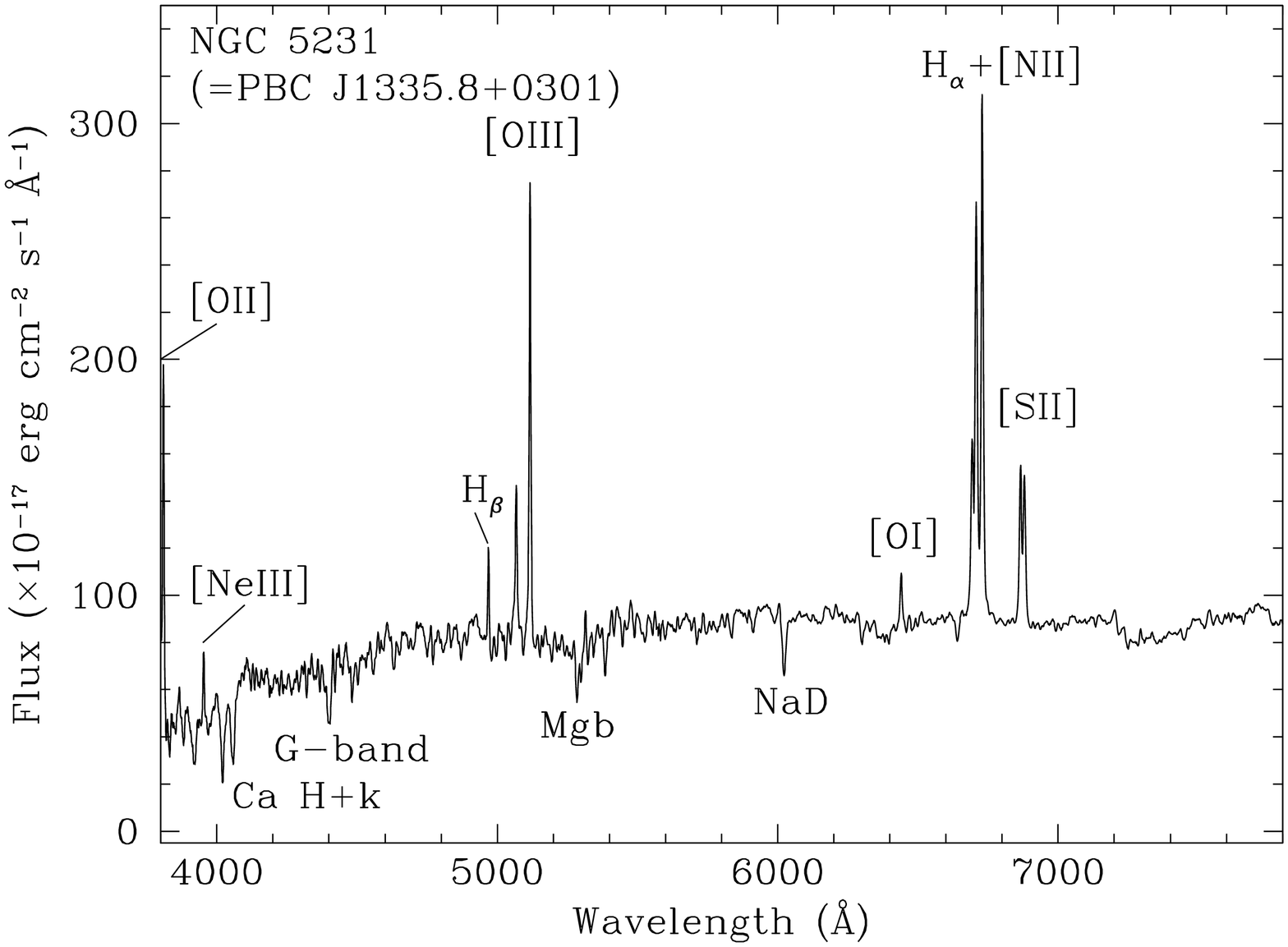,width=6.0cm,angle=270}}}
\centering{\mbox{\psfig{file=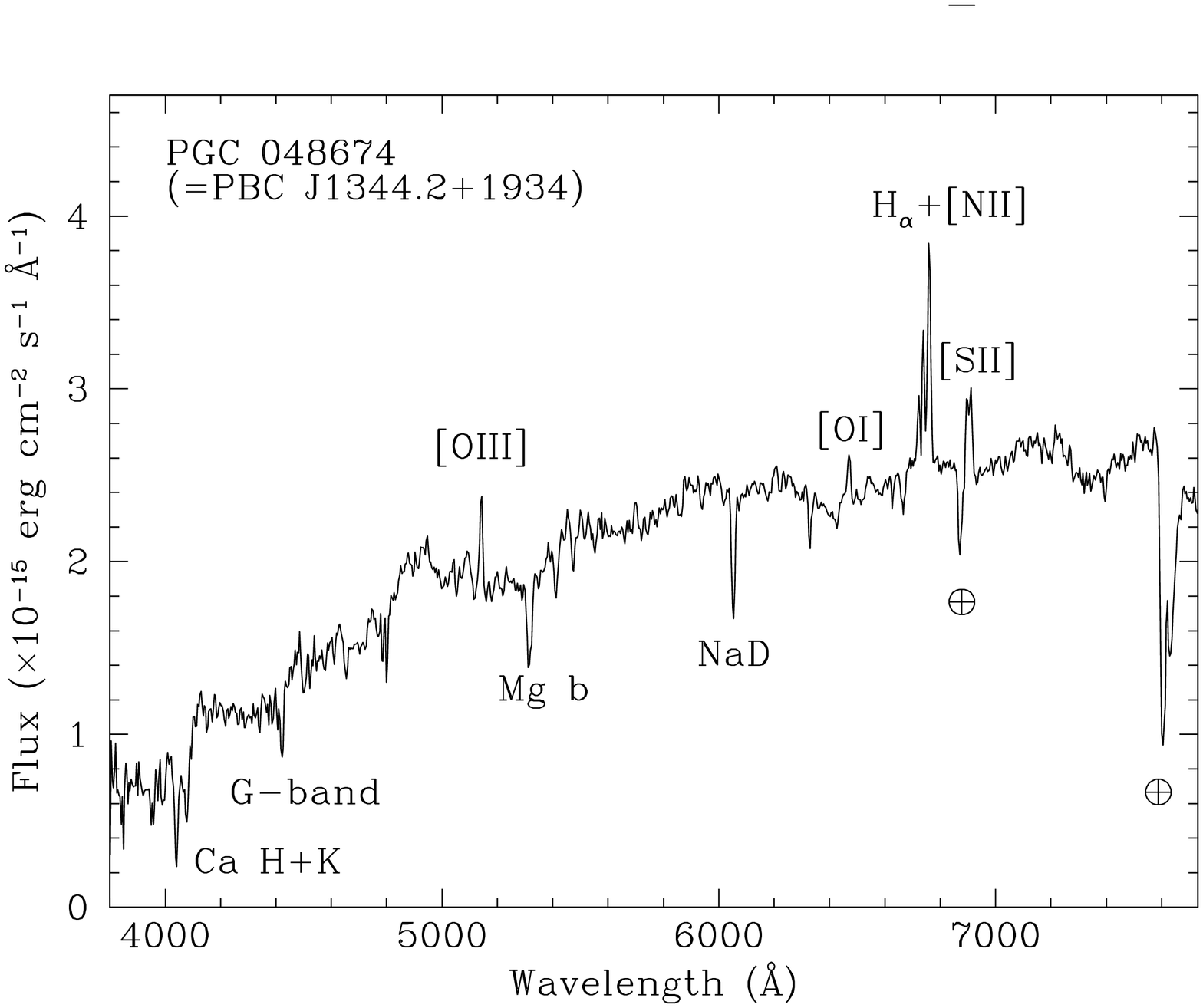,width=6.0cm,angle=270}}}
\centering{\mbox{\psfig{file=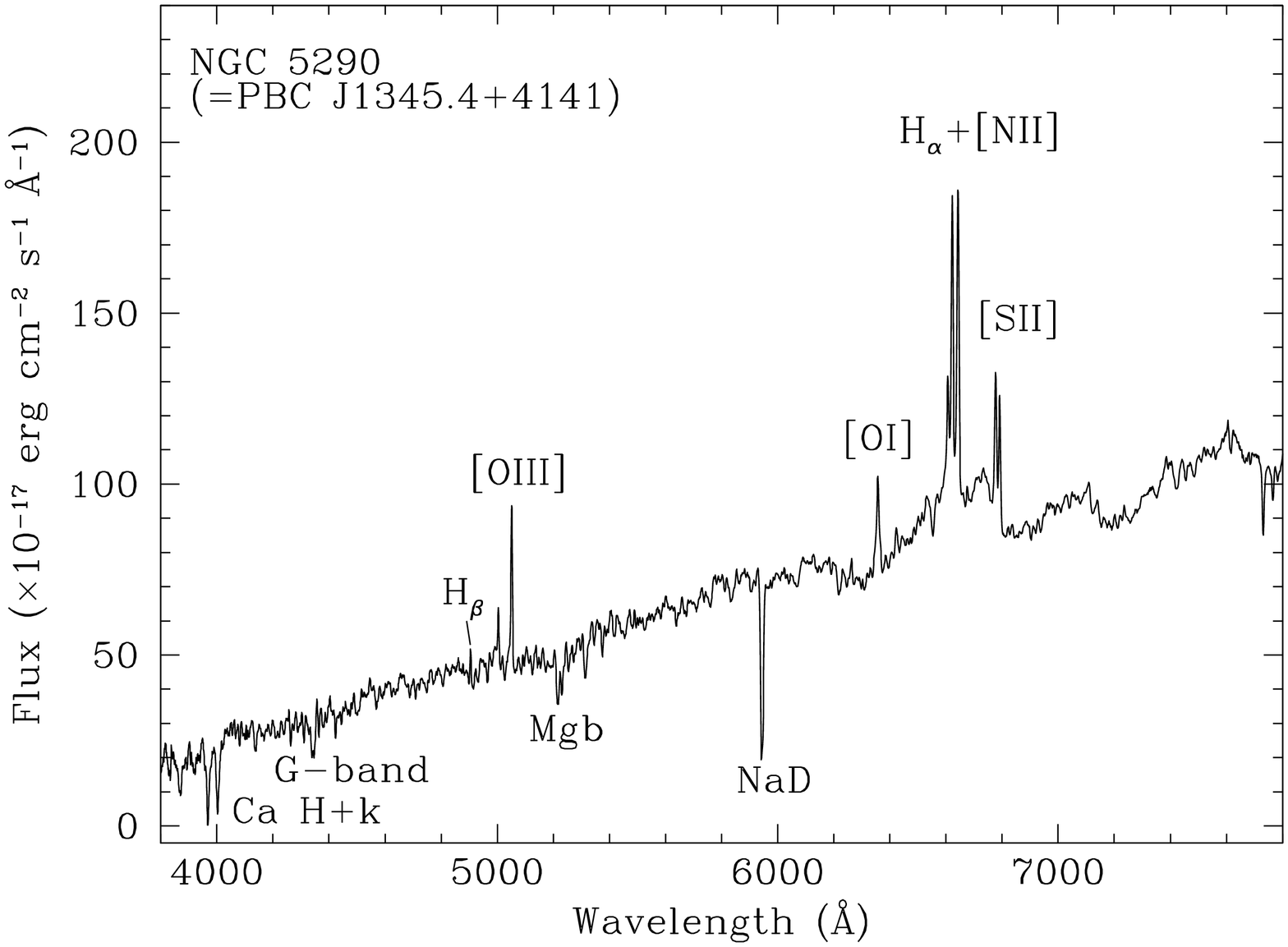,width=6.0cm,angle=270}}}
\centering{\mbox{\psfig{file=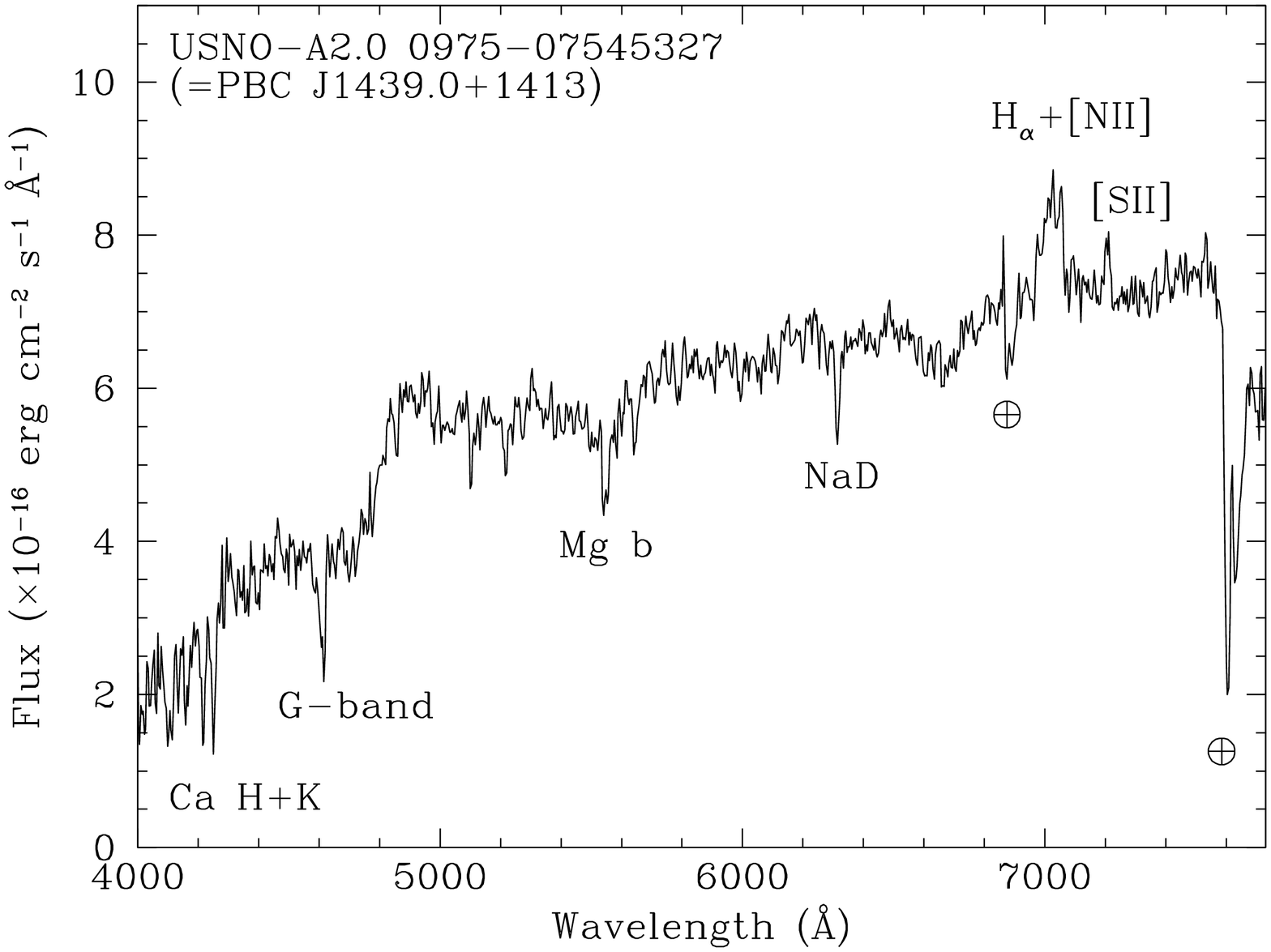,width=6.0cm,angle=270}}}
\centering{\mbox{\psfig{file=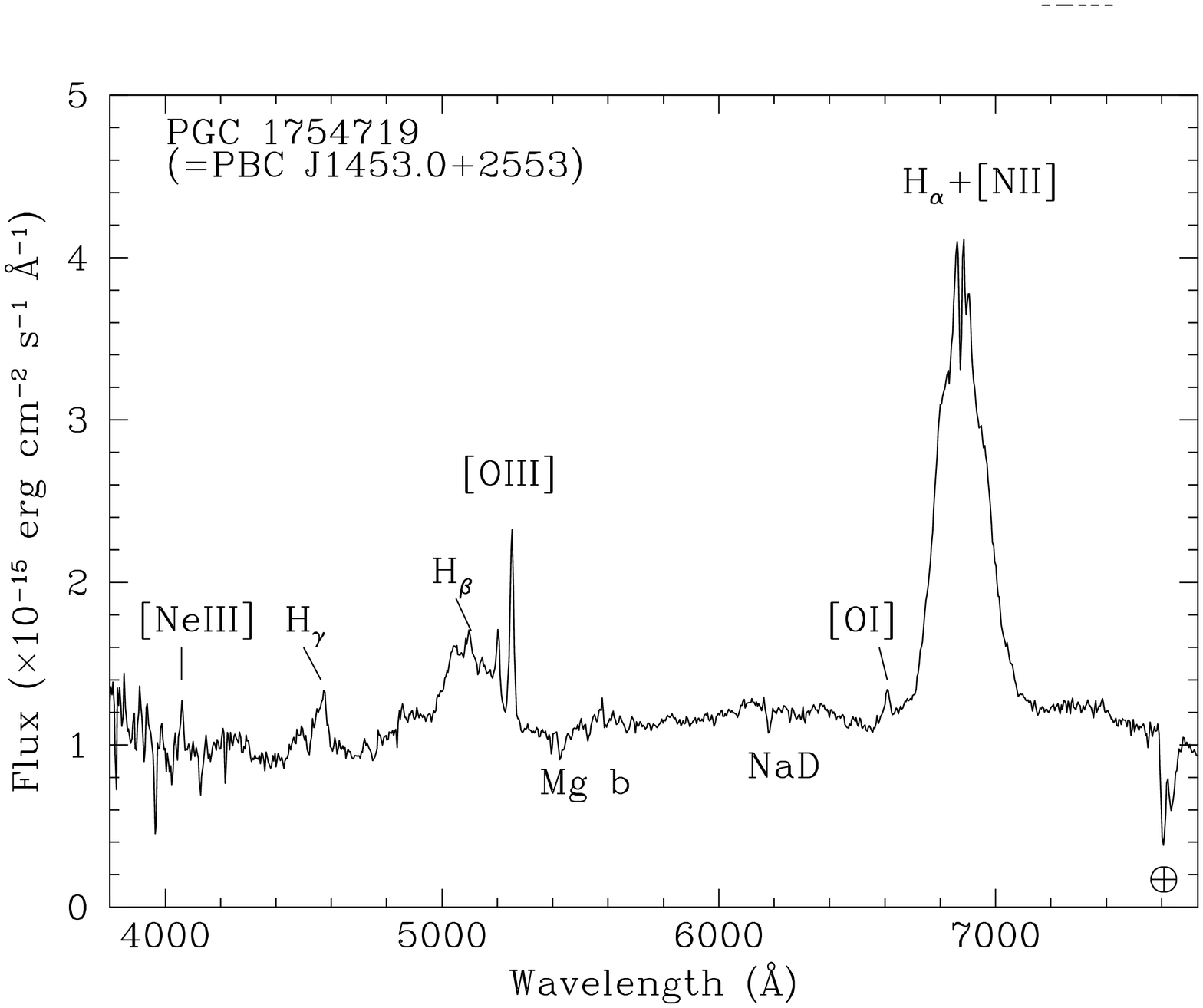,width=6.0cm,angle=270}}}
\centering{\mbox{\psfig{file=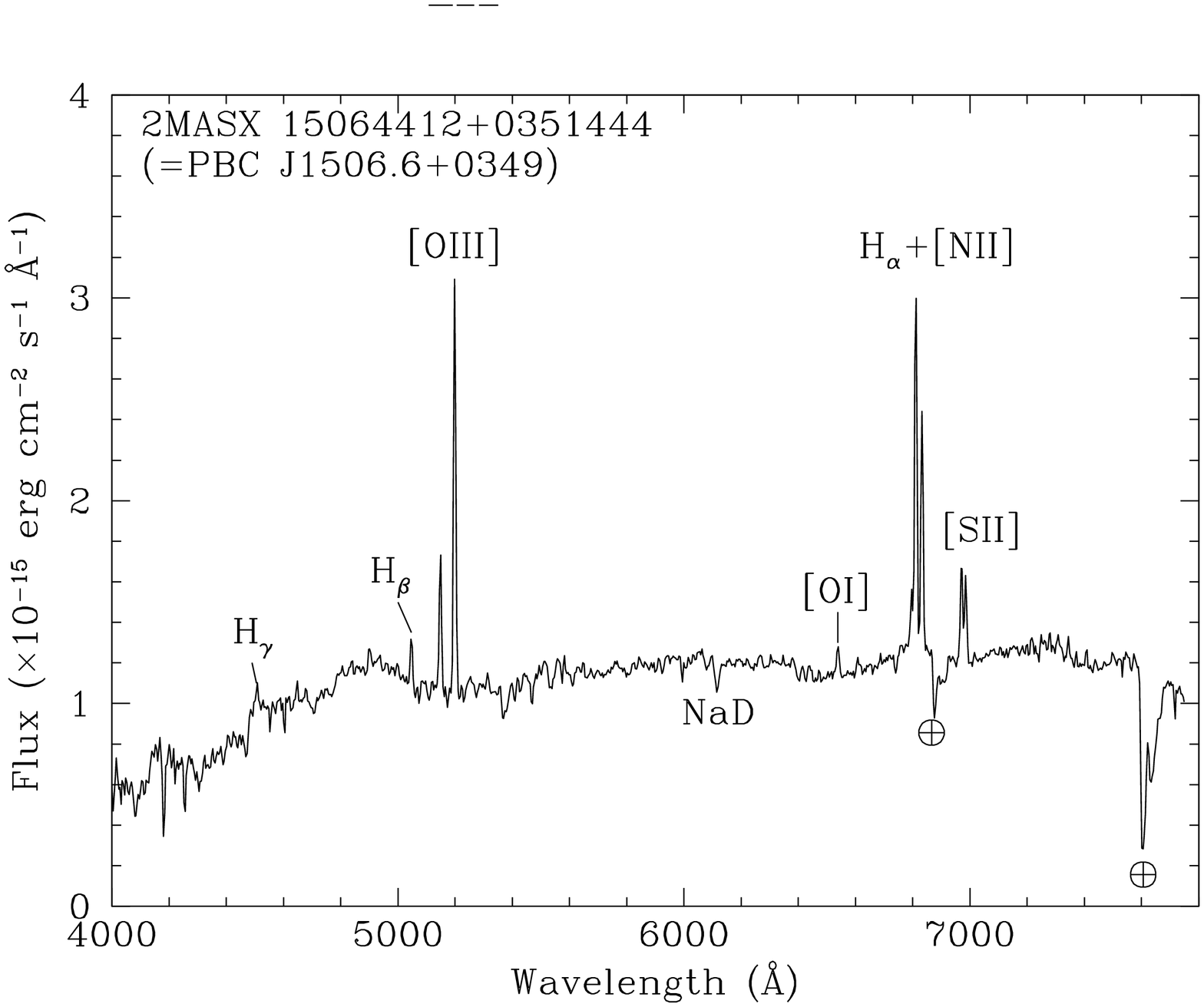,width=6.0cm,angle=270}}}
\centering{\mbox{\psfig{file=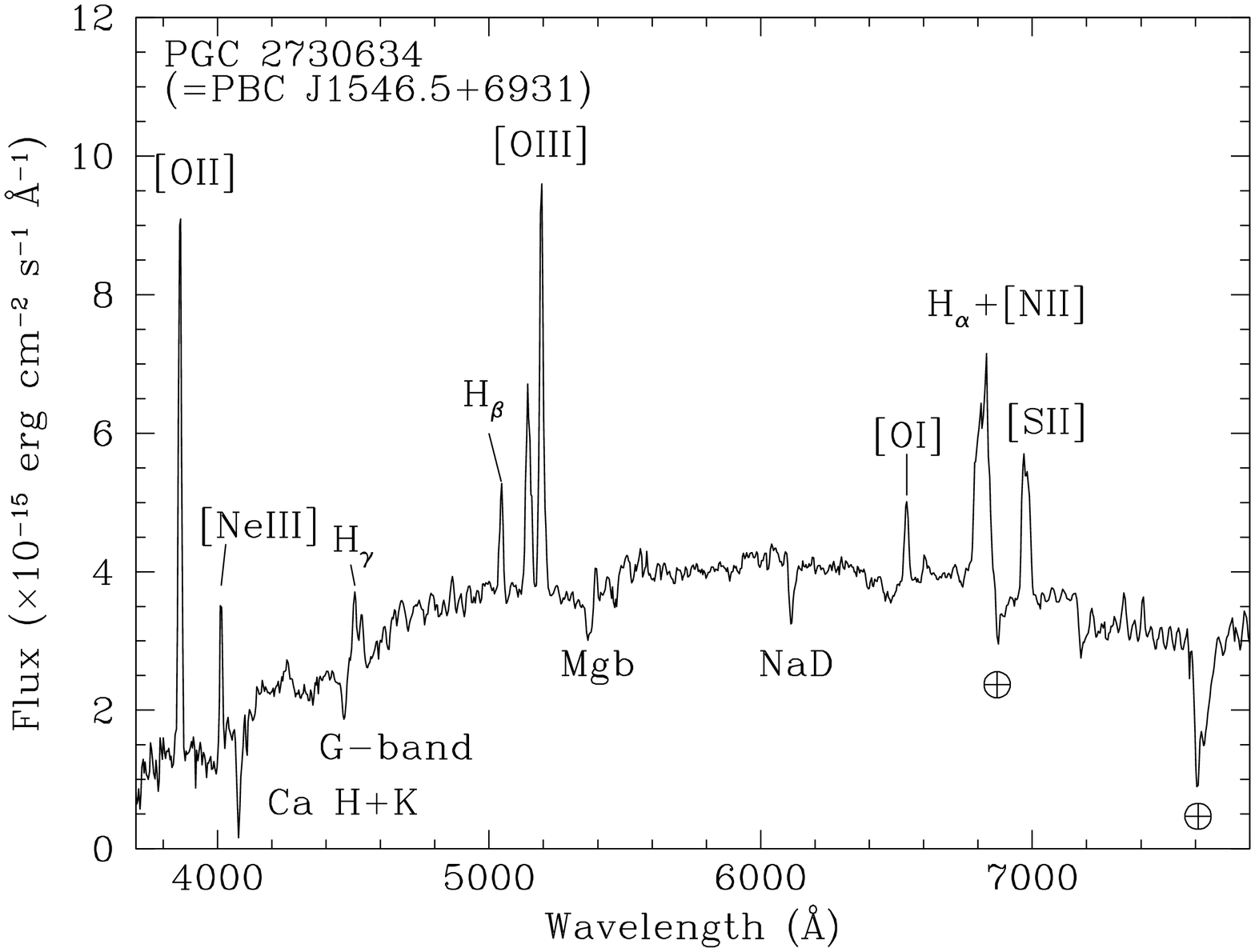,width=6.0cm,angle=270}}}
\centering{\mbox{\psfig{file=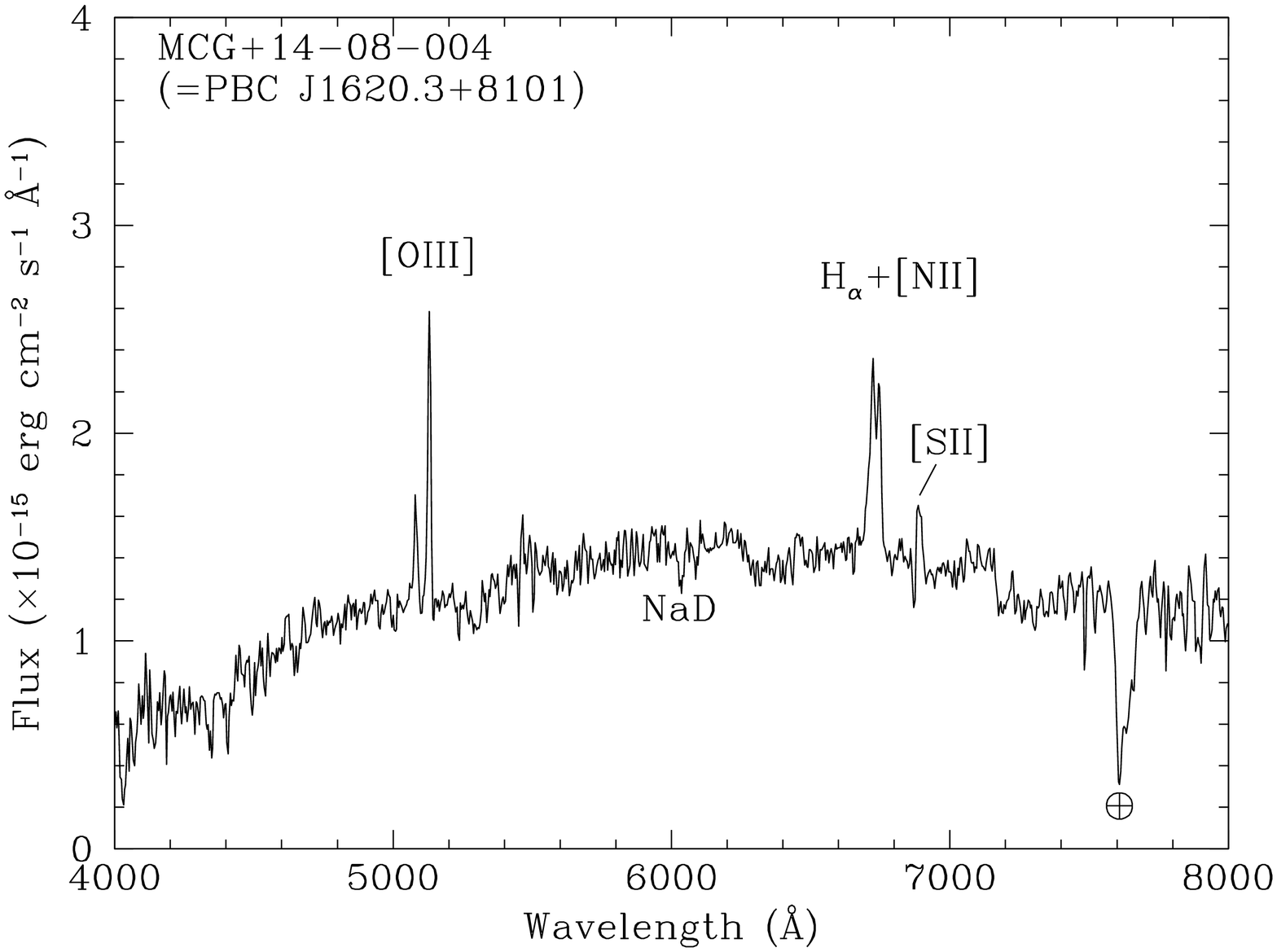,width=6.0cm,angle=270}}}
\centering{\mbox{\psfig{file=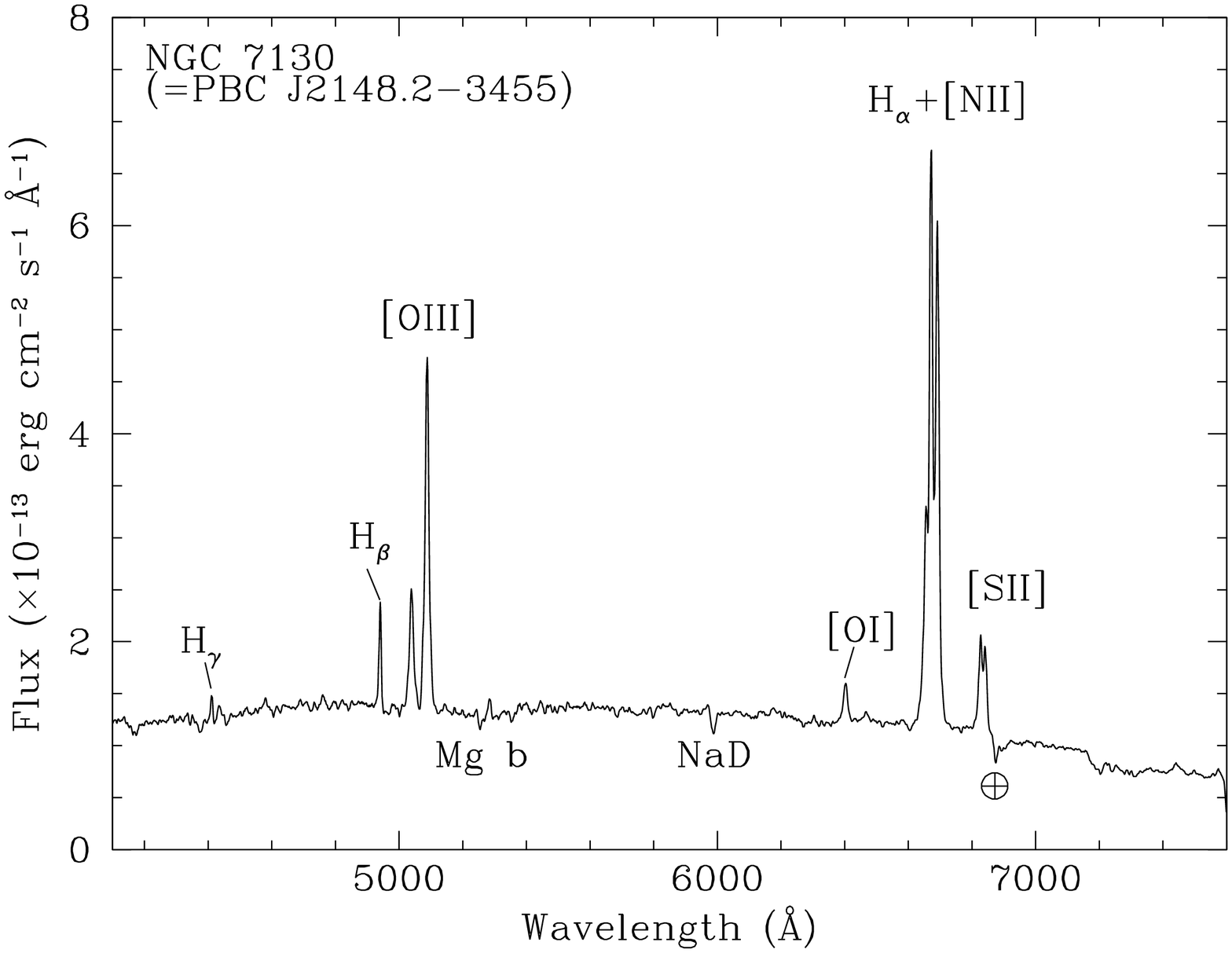,width=6.0cm,angle=270}}}
\centering{\mbox{\psfig{file=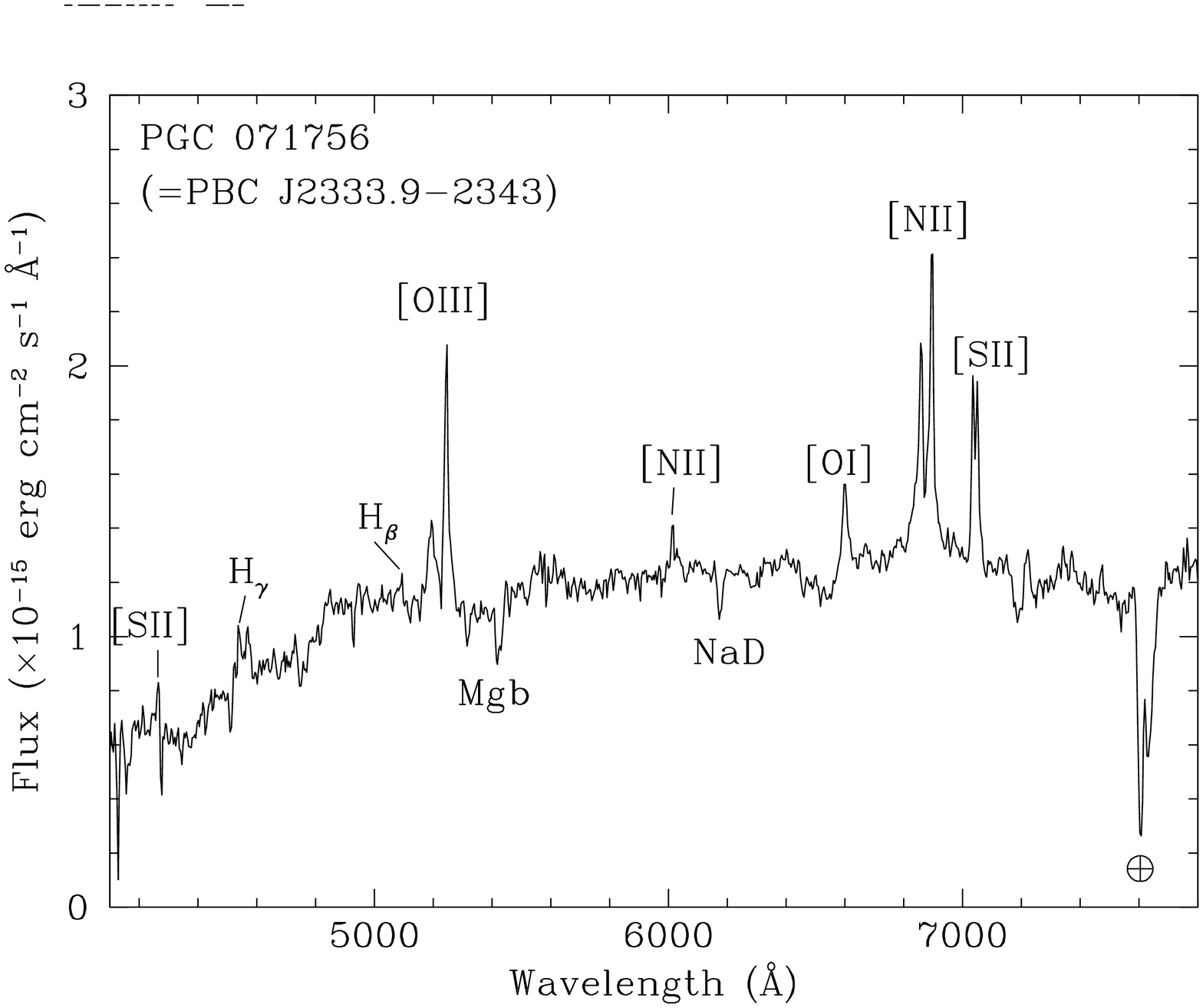,width=6.0cm,angle=270}}}
\centering{\mbox{\psfig{file=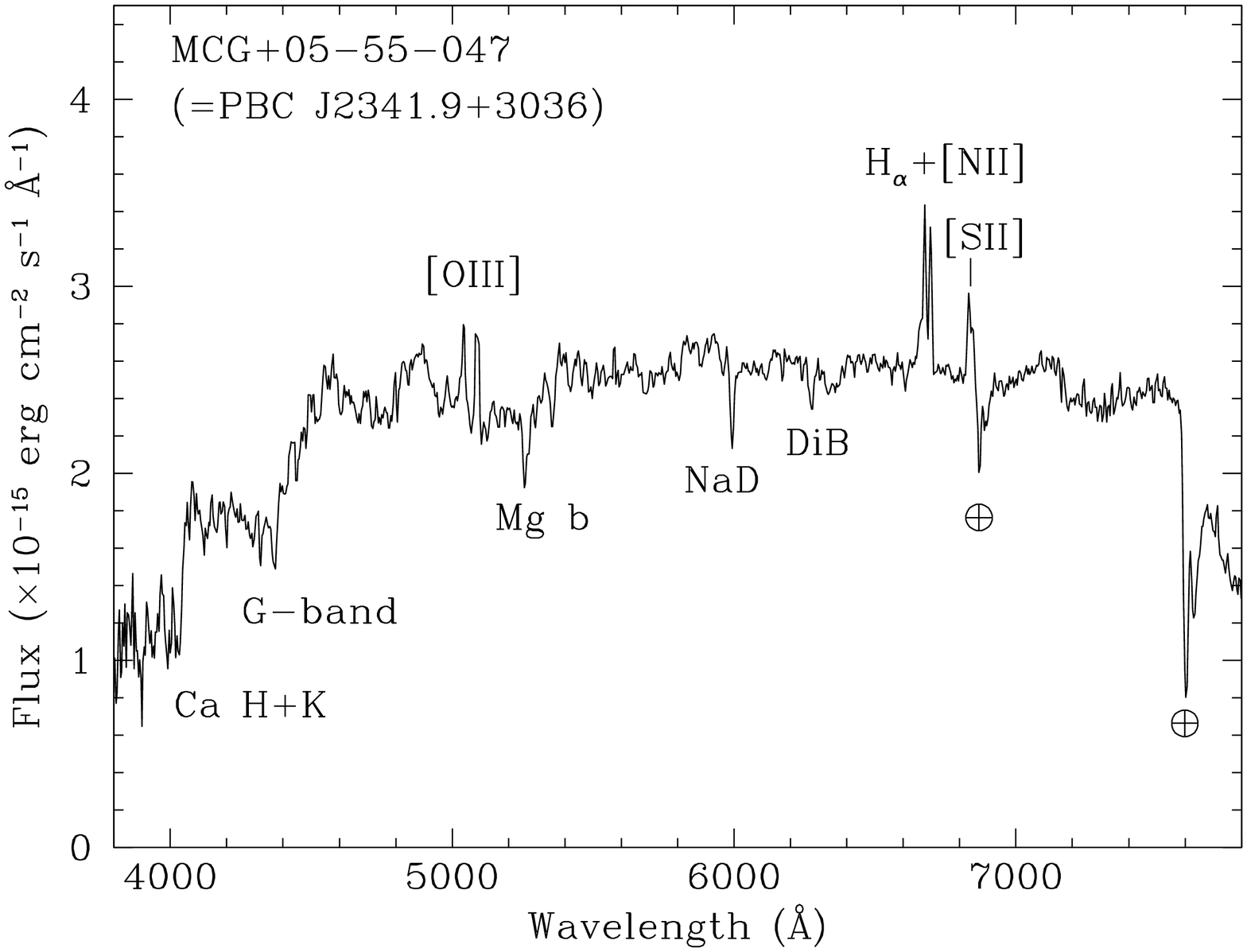,width=6.0cm,angle=270}}}
\caption{Spectra (not corrected for the intervening galactic absorption) of the 
optical counterpart of  PBC J0954.8+3724, PBC J1246.5+5432,
PBC J1335.8+0301, PBC J1344.2+1934, PBC J1345.4+4141, PBC J1439.0+1413, PBC J1453.0+2553, PBC J1506.6+0349, PBC J1546.5+6931, PBC J1620.3+8101, PBC J2148.2$-$3455, PBC J2333.9$-$2343 and PBC J2341.9+3036.
}\label{spectra4}
\end{figure*}

\section{Optical Classification}
In the following we discuss the optical classifications found and highlight
the most interesting/peculiar objects discovered. The {\it B} magnitudes if not otherwise stated, are extracted from the 
LEDA archive (Prugniel et al. 2005) and the {\it R} magnitudes from the USNO-A2.0 catalogue (Monet et al. 2003).

\subsection {Galactic object}
PBC J0826.3$-$7033 is the only source which shows 
 emission lines of the Balmer complex (up to at least H$_{\epsilon}$), as well as of He {\sc i}  and He {\sc ii}, 
consistent with z = 0, indicating that this object lies within our Galaxy (see Fig. \ref{cv}). 
The analysis of all these optical features indicates that this source is a cataclysmic variable (CV, see Tab. \ref{cv}).
The He {\sc ii}$_{\lambda4686}$/H$_{\beta}$ equivalent width (EW) ratio, less than 0.5, and the EW of these two emission lines 
(only the H$_{\beta}$ EW is larger than 10 \AA) point out that this source is likely a non-magnetic CV
(see Warner 1995, and references therein). The H$_{\alpha}$ to H$_{\beta}$ flux ratio is $\sim$1.5 allowing us to consider the absorption
along the line of sight negligible. This is in line with the hydrogen column density value obtained from the X-ray spectral analysis (see below).
The source was previously detected at 
soft X-ray energies being listed for example in the Rosat Bright source catalogue
(Voges et al. 1999); as many other CVs, also PBC J0826.3$-$7033 is located at relatively high  galactic latitudes, i.e. 18 degrees above the galactic plane.

We also estimated its distance to be 90 pc, i.e. relatively close, assuming no extinction along the line of sight.  
The X-ray spectrum is best fitted with a bremsstrahlung model 
(see Tab. \ref{cvx} for more information).
At the estimated distance, the 2-10 keV source luminosity is around 2$\times$10$^{30}$ erg s$^{-1}$,
relatively low compared to the CV so far detected in hard X-rays (Landi et al. 2009).
Finally we estimated the mass of the white dwarf using the Eqs. 5 and 6 of Patterson \& Raymond (1985). Using the bremsstrahlung temperature of our model  
(see Table. \ref{cvx}) and a 0.2-4 keV luminosity of $\sim$ 2.5$\times$10$^{30}$ erg s$^{-1}$, we obtained a value of about 0.4 M$_{\odot}$.

\subsection {Extragalactic objects}

The results of our optical study on extragalactic sources are reported in Tables \ref{agn1} and \ref{agn2}, where we list for each source 
the H$_{\alpha}$, H$_{\beta}$ and [O{\sc iii}] fluxes, the classification, the redshift estimated from the narrow lines, 
the luminosity distance given in Mpc, the galactic color excess and the color excess local to the AGN host.
All the extragalactic optical spectra are displayed in Figures \ref{spectra1} and \ref{spectra4}.
Of the 28 active galaxies found, seven show strong redshifted broad and narrow emission lines typical of Seyfert 1 galaxies,
while the remaining 21 display only strong and redshifted narrow emission lines
typical of Seyfert 2 galaxies (for the subclass classification see Tables \ref{agn1} and \ref{agn2}). 
As reported before, some sources have a preliminary classification and/or redshift in the Palermo 54 months BAT catalogue (Cusumano et al. 2010b), while here we publish for the first time their optical spectra and the corresponding information.
\subsubsection {Redshifts}

Concerning the redshift estimates we confirm the values reported in NED, V\&V13 and Ciroi et al. (2009) for 19 AGN. In one case (PBC J0623.8-3212) we obtain a 
z value (0.035) different {\bf from} the one (0.022) already available, despite the fact that both redshifts were extracted from the same 6dF spectrum. The origin of this discrepancy is not clear.
For the remaining 9 sources we report the redshift obtained by low resolution optical spectra for the first time (see Figs. \ref{spectra1} and \ref{spectra4}).
Redshifts values are in the range 0.008-0.075, which means that our sources are all located in the local Universe. 

All redshifts have been estimated from the [O{\sc iii}] narrow emission line and when these are unavailable, from the forbidden narrow emission lines or absorption features.


\subsubsection {Optical class}
For the first time we also provide the classification of 15 sources in the sample.
For the remaining 14 objects our results only partially (50\%) agree with the classifications listed in the literature. 

For 5 objects (PBC J0100.6-4752, PBC J0356.9-4040, PBC 0503.0+2300, PBC 0814.4+0421 and PBC J1335.8+0301)
we find a different AGN type  than the one already reported. The change is slightly different for the first two sources which are now classified as Seyfert 2 galaxies rather than Seyfert 1.8-1.9 (Baumgartner et al. 2008; Winter et al. 2009); the issue is instead more important for the remaining 3 objects.
PBC 0503.0+2300 and  PBC 0814.4+0421 move from a Seyfert 1 classification (Cusumano et al. 2010a) to a Seyfert 1.5 and to a Narrow Line Seyfert 1 (NLS1)   
respectively. PBC J1335.8+0301 shifts from a type 1 (V\&V13) to a type 2 class.

PBC J2148.2-3455 (named also NGC 7130 or IC 5135) is a known AGN, but with multiple classification.
Phillips et al. (1983), Heisler et al. (1997) and Vaceli et al. (1997) classified it as a Seyfert 2 galaxy, Thuan (1984) and Veilleux et al. (1997)
assigned it to the LINER type, while NED and V\&V13 list it as a Seyfert 1.9.
We confirm the Seyfert 2 nature of PBC J2148.2-3455 and further suggest that the differences in the optical classification may simply reflect 
different contributions from the starburst emission in the observations.

Baldwin et al. (1981) classified PBC J2333.9-2343 (also PKS 2331-240) as a Seyfert 2 galaxy,
Radivich \& Kraus (1971) instead reported  it as a Narrow Line radio galaxy while Andrew et al. (1971) classed it as a QSO, but with no redshift.
Bolton (1975) defined the  object morphology as an elliptical galaxy; as a result NED classified it as a Seyfert of unclear type. 
We are finally been able to classify the object as a Seyfert 2, thus confirming the original indication.

\subsubsection{Peculiar sources and discussion}

Within the sample of type 1 AGN listed in Table \ref{agn1}, we find that one is a Seyfert 1, 5 are AGN of intermediate type (1.2-1.9) and  
one is a NLS1.

In terms of the unified model, intermediate Seyferts have been interpreted as objects in which our line of sight progressively intercepts the obscuring torus starting from its outer edge.
However, this is not the only interpretation as intermediate classifications may be related to other phenomena such as an intrinsically variable 
ionizing continuum: for example,
a source that would normally appear as a Seyfert 1 can be classified as an intermediate type AGN when found in a low flux state (Trippe et al. 2010).
Insights into the properties of our intermediate Seyferts can help in discriminating between the various scenarios (see next section).
 
PBC J0814.4+0421 deserves a  special mention among type 1 AGN. LEDA 023094, which is its optical counterpart with magnitude B = 15.5 and redshift of 0.027, 
displays optical features typical of NLS1 (see Section 3). 
These sources are rare among hard X-ray selected objects 
since their fraction is only 5$\%$ among all AGN and 10$\%$ among type 1 Seyferts (Panessa et al. 2011).  A possible interpretation for the peculiar observational
properties of NLS1 is that these systems are accreting close
to their Eddington limit, implying that, compared to typical Seyfert 1 galaxies, they should host black holes
with smaller masses (M$_{BH}$ $\le$ 10$^7$ M$_{\odot}$). Indeed PBC J0814.4+0421 has the smallest black hole mass among the four objects for 
which this parameter has been estimated in this work; 
the value obtained is also compatible with those of other hard X-ray selected NLS1 (Panessa et al. 2011).
Interestingly,  PBC J1453.0+2553, the only object classified as a pure Seyfert 1 has by far the biggest black hole mass observed among our small sample 
of type 1 AGN.

A large fraction of our extragalactic objects belongs to the type 2 AGN  class; this is not unexpected, as hard X-ray surveys are very efficient
in discovering this type of galaxies. Also among type 2 AGN there are a few interesting cases, 
such as LINERs.

PBC J0041.6+2534 and PBC J1344.2+1934, for example, are located in 
an intermediate region between Seyfert 2 and LINERs of the diagnostic diagrams (Ho et al. 1993; Kauffmann et al. 2003);
because of this proximity to Seyfert 2 galaxies, both are treated here as type 2 AGN. PBC J0353.5+3713 is instead a pure LINER, but since displays only narrow emission lines in the optical spectrum
it is also considered as a type 2 AGN. Interestingly, all 3 objects with LINER signatures are absorbed in X-rays, thus confirming their similarity with type 2 AGN.

As a final remark,  we note that all our AGN have X-ray spectra typical of their class; that is, a simple power law (intrinsically absorbed or not) 
plus in many cases
an extra soft component which can be parameterized by a second power law (having the same photon index of the primary component) or  by a black body model. 
Only in 3 objects 
we detect emission lines compatible with neutral (in two cases) and ionized (only in one case) iron\footnote{Because of the XRT sensibility we cannot reach the 6-7 keV to detect the iron features in all data.}. We will not comment on these spectra further, but we will use 
the information on the intrinsic column density to compare in the following sections the optical versus X-ray classification and to discuss 
the optical (dust) versus X-ray (gas) absorption.

\section{X-ray versus optical classification and absorption}
The unification scheme states that every AGN is intrinsically the same object, an accreting supermassive black hole surrounded by an obscuring torus:
depending on how the observer views the central engine we classify AGN either as type 1 (where we see directly to the nucleus and hence both broad and narrow line regions are visible)  or as type 2 
(where we
see the nucleus through the torus which hides the broad line region but not the narrow line one). Because the torus is made of dust and gas, we expect that type 2 AGN are absorbed in X-ray and optical wavelengths and type 1 are not.
The X-ray absorption is directly measured by the X-ray column density while the optical one is estimated by means of the  colour excess.  

\begin{figure}[th!]
\hspace{-.1cm}
\centering{\mbox{\psfig{file=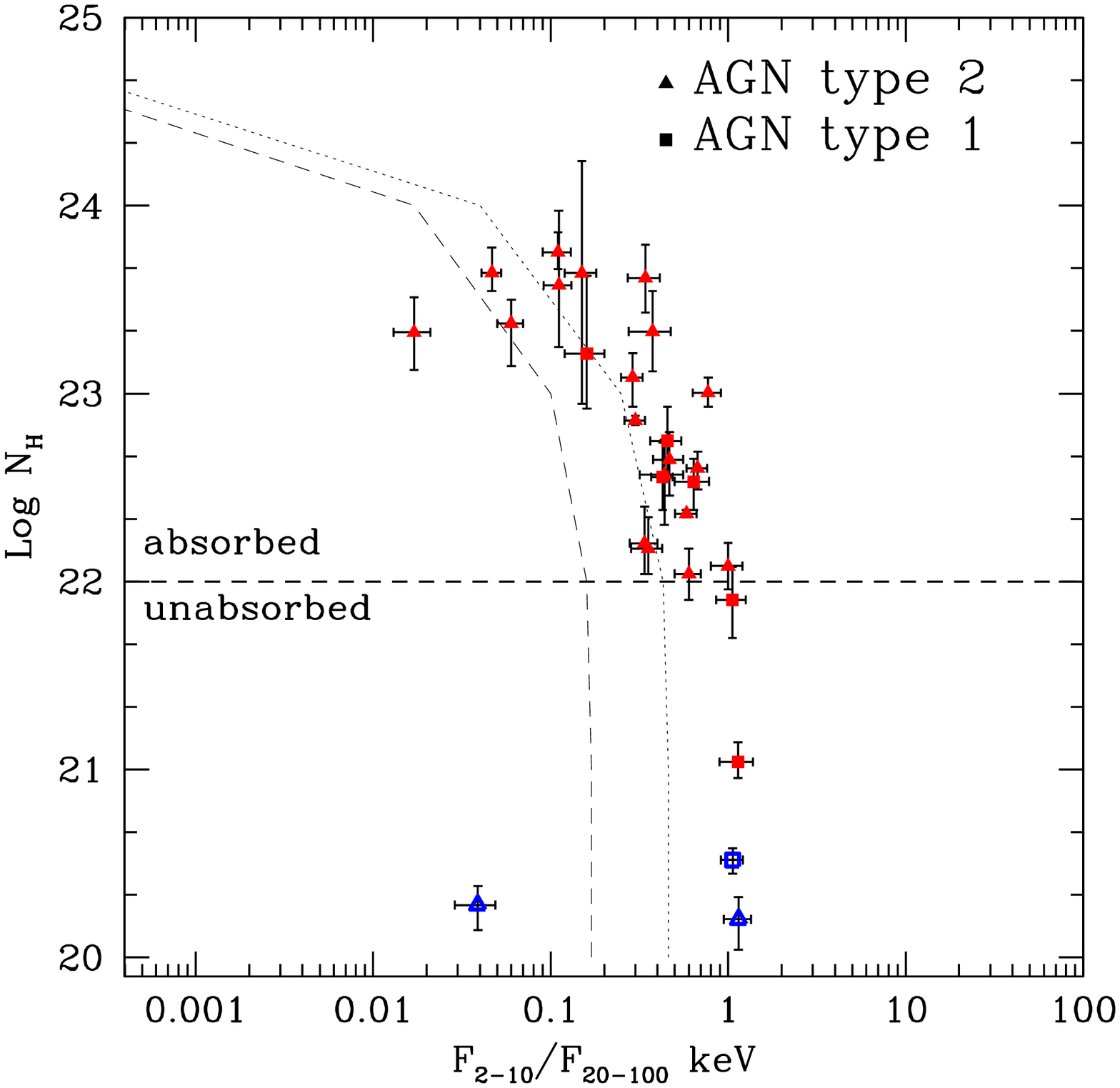,width=9cm,angle=0}}}
\centering{\mbox{\psfig{file=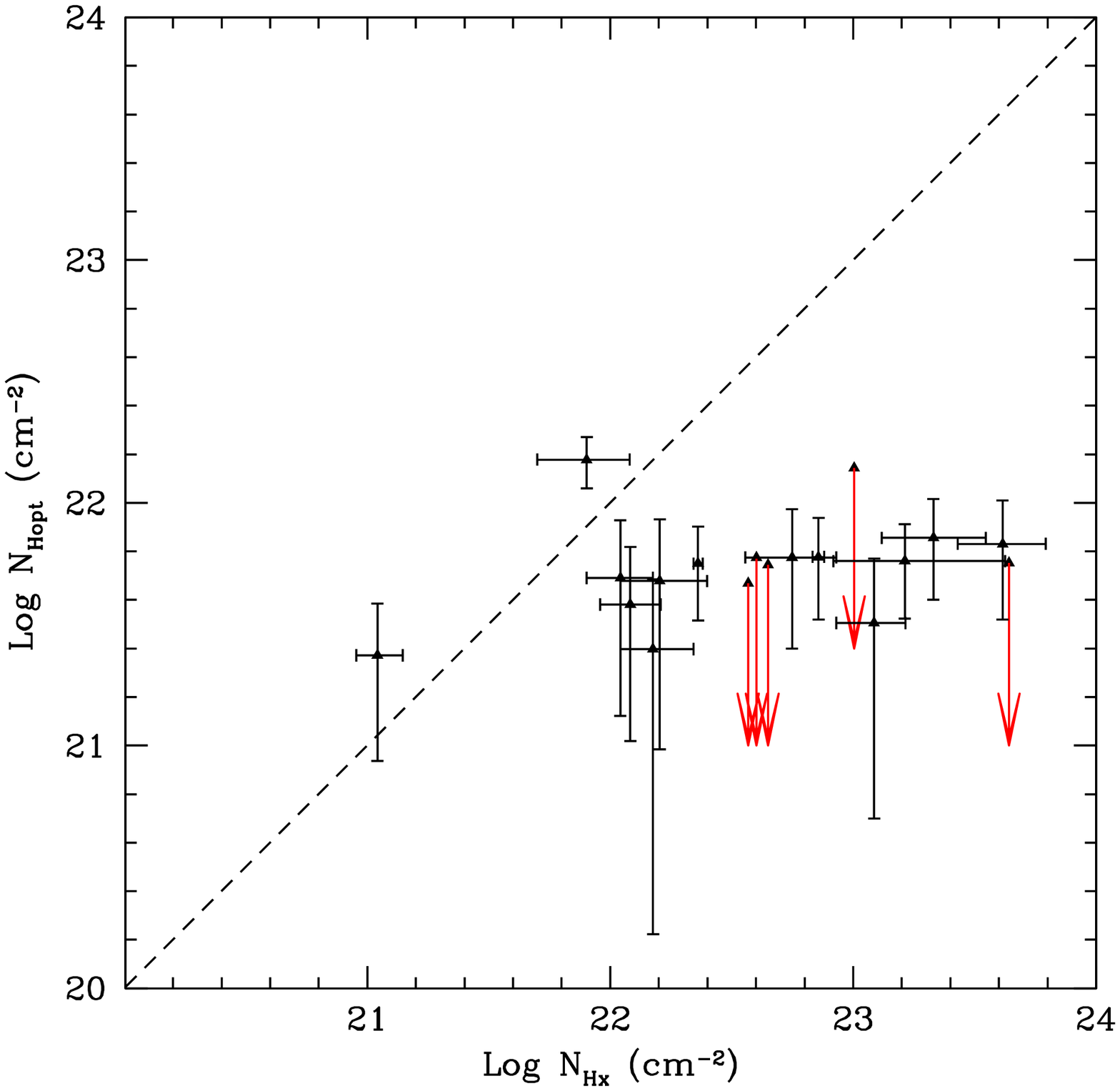,width=9cm,angle=0}}}
\caption{{\it Top panel:} F$_{2-10 keV}$/ F$_{20-100 keV}$ flux ratio of our sample. Lines correspond to expected values for an absorbed power law
with photon index 1.5 (dotted) and 1.9 (dashed); see text for details. The blue empty symbols indicate sources with galactic absorption only, while red filled ones correspond to sources with an intrinsic one also.
{\it Bottom panel:} X-ray column density versus the optical one computed from the E(B-V) assuming the Galactic extinction law of Cardelli et al. (1989). Triangles are Seyfert 2, squares are Seyfert 1 objects.
Points marked with red arrows are upper limits.
}\label{ratio}
\end{figure}

It is however important to note that, sometimes, heavily absorbed objects also known as  Compton thick AGN,  appear `unabsorbed' in X-rays due to the poor
statistical quality of the X-ray data and/or the lack of high energy information. To recognize such objects we can use the diagnostic diagram of Malizia et al. (2007), 
which plots the X-ray absorption as a function of the source  flux  ratio F$_{(2-10)keV}$/F$_{(20-100)keV}$: for our sample this is done 
in  Fig. \ref{ratio}, where N$_H$ is for most objects  the intrinsic value  measured (see Table \ref{fit}) and, for sources PBC J1453.0+2553, PBC J2148.2$-$3455 and PBC J2333.9$-$2343,
the galactic one which is taken here as an upper limit to the X-ray absorption. 
A clear trend of decreasing flux ratios as the absorption 
increases is expected and is due to the fact that the 2-10 keV flux is progressively depressed as the
absorption becomes stronger. Indeed, the two lines shown in the figure  describe how the flux ratio is expected to change as a function of N$_H$ 
in the case of objects characterized by  an absorbed power law having a photon index of 
of 1.5 and 1.9, respectively. It is evident that most of our sources follow the expected trend with the most absorbed  AGN showing progressively 
lower F$_{(2-10)keV}$/F$_{(20-100)keV}$ values. Assuming as a dividing line between absorbed and unabsorbed AGN a column density of 10$^{22}$ cm$^{-2}$ 
(sufficient to hide the broad line region of an active  nucleus), we note that most of our  AGN are above this line or very close to it, as expected 
given the high percentage of type 2 AGN in our sample. Indeed
PBC J0503+2300 and PBC J1453.0+2553, a type 1.5 and a type 1 Seyfert respectively, are both well below the line, while 2 out of 3 Seyfert 1.9  galaxies in our sample have a column density 
above 10$^{22}$ cm$^{-2}$,
i.e. compatible with the idea that these objects are looked through the edge of the torus;  the third one, PBC J1546.5+6931,  
has a column density just below the dividing line between absorbed and unabsorbed objects. 

PBC J0814.4+0421, the only NLS1 in the sample,  is absorbed, 
but this is not unusual, as the presence of strong, partial and/or stratified absorption is one of the two competing models used to explain the 
complex X-ray spectra of this class of AGN (Panessa et al. 2011).
 
The only type 1 AGN for which absorption is totally unexpected is PBC J0543.6-2738, classified as a type 1.2 AGN, but displaying a column density in the range
(1.4-5.5) $\times$ 10$^{22}$ cm$^{-2}$; it is possible that in this source the gas responsible for the X-ray absorption is highly ionized,
instead of being neutral, in which case the accompanying
dust would sublimate, yielding a much smaller dust-to-gas
ratio and resulting in a reduced optical extinction and consequently in an early type 1 classification.
From this perspective, the obscuring material would not be related to the toroidal structure as assumed in the unified theory, but rather to other types of absorption, possibly 
gas in an outflow from the central nucleus (Winter et al. 2011). Unfortunately, our X-ray spectrum does not have a signal-to-noise ratio
high enough to distinguish between ionized and  neutral
gas models; clearly  PBC J0543.6-2738 deserves a much more in-depth X-ray study with satellites such as \emph{XMM-Newton} or Suzaku.
 
\begin{table}[h!]
\caption[]{BLR gas velocities and 
central black hole masses for 4
Seyfert 1 AGNs listed in this paper.}
\label{blr}
\begin{center}
\begin{tabular}{lcc}
\noalign{\smallskip}
\hline
\hline
\noalign{\smallskip}
\multicolumn{1}{c}{Object} & $v_{\rm BLR}$ & $M_{\rm BH}$ \\
\multicolumn{1}{c}{}& (km s$^{-1}$)&(10$^7$ $M_\odot$)\\
\noalign{\smallskip}
\hline
\noalign{\smallskip}

PBC J0503.0+2300   & 3600 & 5.2 \\
PBC J0543.6$-$2738 & 4500 & 7.1  \\
PBC J0814.4+0421 &1800  & 1.1 \\
PBC J1453.0+2553 & 10800  & 43  \\
\noalign{\smallskip} 
\hline
\hline
\end{tabular}
\end{center}
\end{table}

On the other hand, we find two objects which are classified as 
type 2 Seyferts in optical, but show  no absorption in X-rays: PBC J2148.2$-$3455 and PBC J2333.9-2343. As anticipated above, it is possible that 
these 2 AGN are Compton thick sources not recognized as such due to the 
poor quality of the X-ray data; but while this may be the case of the first source, 
it is certainly not true for the second one, because of the following reasons.

PBC J2148.2$-$3455  is well studied in X-rays and previous observations strongly indicate that this is indeed a Compton thick AGN. By means of Chandra data, 
Levenson et al. (2005) showed that the active nucleus is probably buried beneath a column
density  N$_{Hx}$ $\ge$ 10$^{24}$  cm$^{-2}$ as indicated by the prominent Fe 
K${\alpha}$ emission line which has an equivalent width larger than 1 keV; also the F$_{2-10}$ keV/F[O{\sc iii}] ratio, often used as an alternative way to pinpoint 
heavily absorbed Seyfert 2 galaxies, 
is sufficiently small (0.04) to classify the source as a Compton thick object (Bassani et al. 1999). 
PBC J2333.9-2343, on the other hand is quite atypical: it lies in the region of type 1 AGN (see Fig. \ref{ratio}, top panel), 
has a good quality spectrum which provides no indication for the presence of an iron line and has a
F$_{2-10}$ keV/F[O{\sc iii}] ratio of $\sim$ 1, again similar to type 1 Seyferts (Bassani et al. 1999).
The source is peculiar in many other ways: it is listed in the Roma BZCAT as a blazar of unknown type (Massaro et al. 2009), 
it is a flat spectrum radio source (Healey et al. 2007), it shows a jet feature in a VLBI 8.4 GHz image, it is variable in radio (Ojha et al. 2004) and also in X-rays according to a quick look analysis of all publicly available Swift/XRT observations and finally 
it is polarized at radio frequencies 
(Ricci et al. 2004). This source looks like a nearby blazar, but it has the optical spectral appearance of a type 2 Seyfert, which is quite unusual since flat spectrum radio quasars (one of the two types of blazars) are generally broad line AGN.
This is an object that certainly deserves further investigation in X-rays, but also in other wavebands to confirm the above peculiarities.\\

Summarizing, we find that the optical versus X-ray classifications for most of our sources broadly fit with the AGN unified scheme except for a few peculiar
objects: PBC J0543.6-2738, which is a type 1 AGN showing some absorption possibly due to outflowing gas, and PBC J2333.9-2343 which is instead a type 2 AGN displaying no absorption and with properties similar to blazars. 

In Fig. \ref{ratio} bottom panel, we plot the X-ray column density  versus the optical one (with relative uncertainties), measured from the intrinsic colour index E(B-V) using the following formula: 
N$_{Hopt}$= 3.1 E(B-V) $\times$ 1.79 $\times$ 10$^{21}$  cm$^{-2}$ (Predehl \& Schmitt 1995; Rieke \& Lebofski 1985). 
For 5 sources we decided to take 90\% upper limits (marked with red arrows, in Fig. \ref{ratio}) because the errors were larger than the N$_{Hopt}$ values. 
Within the unified theory, the X-rays absorption is associated with the gas in the  torus which is confined to a region smaller than 
the Narrow Line Region (NLR);
 the optical extinction, on the other hand, may come from dust either associated with the torus (internal reddening) or to larger scale structures 
like lanes, bars or something else (external reddening). 
As evident from the figure the majority of our sources  has an X-ray column density higher that the optical one. 
Since most of our objects are AGN of type 2, the optical reddening is related to the NLR and hence most likely associated with internal rather 
than external reddening, i.e. the torus; in this case the bottom panel in Fig. \ref{ratio} is simply telling us that, with this structure, gas absorption is higher than dust obscuration.
This effect was already noticed by Maiolino et al. (2001) who 
suggested as a viable explanation that in AGN the dust-to-gas ratio is much lower than the galactic one or that in the inner part of the obscuring torus 
the dust is sublimated by the strong UV radiation field. Another interesting possibility put forward by the same authors is that 
the dust extinction curve is much flatter than the standard galactic one, for example as a result of the growth of larger dust grains. Only 3 of our objects
are located above the 1-1 line implying an optical/(dust) extinction similar or slightly higher than the X-ray/(gas) absorption. Two of these objects are type 1 AGN 
and so this is to be expected,
as our line of sight to their central nucleus does not intercept the torus; the third object is PBC J1345.4+4141, the only type 1.9 AGN of our sample which has an X-ray column 
density close to but below
10$^{22}$  cm$^{-2}$. In this case, it is possible that the optical reddening and the X-ray absorption  are unrelated to the torus, but may come from other
larger scale structures.  It is interesting to note that NGC 5290, the optical counterpart of PBC J1345.4+4141, forms a pair of interacting galaxies with NGC 5289 
(van Driel et al. 2001); the latter also shows a bright bulge, partially hidden by a dark lane and asymmetric absorption (see NED notes).
In other words there is plenty of reddening  on large scales to explain the observed properties.  It is also possible that in this source 
the BLR optical continuum has temporarily diminished leaving visible only a very weak broad H$\beta$ line and causing the classification as a type 1.9 
even if the nucleus is totally unobscured by dust (and also by gas given the not so high column density). This is another object for which further observations are clearly encouraged especially at optical/infrared wavelengths.

\section{Conclusions}

In this work, we have either given for the first time, or confirmed, or corrected, the optical spectroscopic identification of 29  
sources belonging to the Palermo 39 months {\it Swift}/BAT catalogue (Cusumano et al. 2010a). 
This was achieved through a multisite observational campaign in Europe, South Africa and Central America.

We found that our sample is composed of 28 AGN (7 of type 1 and 21 of type 2), with redshifts between 0.008 and 0.075, and 1 CV.
Among the extragalactic sources we found some peculiar objects, such as 3 AGN showing LINER features and 1 with the properties of NLS1. 
For four type 1 AGN we have estimated the BLR size, velocity and the central black hole mass.
We have also performed an X-ray spectral analysis of the entire sample and found that overall our sources display X-ray spectra typical of their optical class.
More specifically, we have compared the optical versus X-ray classification of our galaxies, in order to test the AGN unified theory.
We find a generally good match between optical class and X-ray absorption, thus providing evidence for the validity 
of the  unified scheme. However, in a few sources there is a clear discrepancy between optical and X-ray classification:
PBC J0543.6-2738 is a Seyfert 1.2 displaying mild X-ray absorption, possibly due to outflowing gas; PBC J1345.4+4141 is instead a Seyfert 1.9 
showing no absorption although its optical class may be related to reddening occurring on large scale structures or  due to a  low optical ionization state.
More convincingly  outside the unified scheme is PBC J2333.9-2343 which is a Seyfert 2 without intrinsic X-ray column density; this source has many features which
make it very similar to broad line blazars and yet has only narrow lines in its optical spectrum.
Another Seyfert 2 displaying no absorption is  PBC J2148.2$-$3455, but through the use of our  diagnostic diagram and information gathered in the literature
we conclude  that this source is a Compton thick or heavily absorbed AGN, which is  therefore compatible with its optical class.
We also compared the X-ray gas absorption with the optical dust reddening for the AGN sample: we find that for most of our sources, specifically those of 
type 1.9-2, the former is higher than the latter, confirming early results by Maiolino et al. (2001); possibly this is due to the properties of dust 
in the circumnuclear obscuring torus of the AGN.

As a final remark, we would like to stress the importance of combining optical with X-ray spectroscopy  for hard X-ray selected objects: using 
information in both wavebands is not only possible to increase the number of source identification and classification, but also  to perform 
statistically significant population studies, to understand the physical processes that occurring in these objects and to study the AGN unified model.

\begin{acknowledgements}

We thank Dr. Domitilla de Martino for useful discussions and the referee for comments which helped us to improve the quality of this paper.
We also thank Silvia Galleti for Service Mode observations at the Loiano 
telescope; Antonio De Blasi and Ivan Bruni for night assistance at the 
Loiano telescope. We also thanks Claudia Reyes for night assistance at the ESO NTT telescope.
This work is based on observations obtained with XMM-Newton, an ESA science mission with instruments and contributions directly funded by ESA Member States and NASA.
We also acknowledge the use of public data from the {\it Swift} data archive.
This research has made use of the ASI Science Data Center Multimission 
Archive, of the NASA Astrophysics Data System Abstract Service, 
the NASA/IPAC Extragalactic Database (NED), of the NASA/IPAC Infrared 
Science Archive, which are operated by the Jet Propulsion Laboratory, 
California Institute of Technology, under contract with the National 
Aeronautics and Space Administration and of data obtained from the High Energy 
Astrophysics Science Archive Research Center (HEASARC), provided by NASA's GSFC.
This publication made use of data products from the Two Micron All 
Sky Survey (2MASS), which is a joint project of the University of 
Massachusetts and the Infrared Processing and Analysis Center/California 
Institute of Technology, funded by the National Aeronautics and Space 
Administration and the National Science Foundation.
This research has also made use of data extracted from the 6dF 
Galaxy Survey and the Sloan Digitized Sky Survey archives;
the SIMBAD database operated at CDS, Strasbourg, 
France, and of the HyperLeda catalogue operated at the Observatoire de 
Lyon, France.
The authors acknowledge the ASI and INAF financial support via grants No. I/033/10/0, I/009/10/0;
P.P. is supported by the INTEGRAL ASI-INAF grant No. 033/1070.
L.M. is supported by the University of Padua through grant No. 
CPDR061795/06. G.G. is supported by FONDECYT 1085267.
V.C. is supported by the CONACyT research grants 54480
and 15149 (M\'exico).
D.M. is supported by the Basal CATA PFB 06/09, and FONDAP Center for 
Astrophysics grant No. 15010003.

\end{acknowledgements}


\begin{thebibliography}{}


\bibitem{} Adelman-McCarthy, J.k., Ag\"ueros, M.A., Allam, S. S., et al. 2007, ApJS, 172, 634




\bibitem{} 	Andrew, B.H., Harvey, G.A., Medd, W.J., 1971, ApL, 9, 151

\bibitem{} 	Baldwin, J.A., Wampler, E.J., Burbidge, E.M., 1981, ApJ, 243, 76

\bibitem{} Barthelmy, S.D. 2004, Proceedings of the SPIE, 5165, 175

\bibitem{bas99} Bassani, L., Dadina, M., Maiolino, R., et al. 1999, ApJS, 121, 473

\bibitem{} Baumgartner, W.H., Tueller, J., Mushotzky, R.F., et al. 2008, ATel 1794







\bibitem{} Bolton, J.G., Shimmins, A.J., Wall, J.V., 1975, AuJPA, 34, 1

\bibitem{} Burrows, D.N., Hill, J.E., Nousek, J.A., et al. 2004, Proc. SPIE, 5165, 201






	
\bibitem{} Cardelli, J.A., Clayton, G.C., \& Mathis, J.S. 1989, ApJ, 345,
       245

\bibitem{} Ciroi, S. et al. 2009, Atel 1985



\bibitem{} Cusumano, G., La Paola, V., Segreto, A., et al. 2010a, A\&A, 510, 48

\bibitem{} Cusumano, G., La Paola, V., Segreto, A., et al. 2010b, A\&A, 524, 64

\bibitem[]{} Dickey, J. M., \& Lockman, F. J. 1990, ARA\&A, 28, 215

\bibitem{} Doyle, M.T., Drinkwater, M.J., Rohde, D.J., et al. 2005, MNRAS, 361, 34





\bibitem{} Gehrels, N., Chincarini, G., Giommi, P., et al. 2004, ApJ, 611, 
	1005
	
	



\bibitem{} Gonz\'alez-Mart\'in, O., Masegosa, J., Marquez, I., et al. 2009 , ApJ, 704, 1570





\bibitem{} Healey, S.E., Romani, R.W., Taylor, G.B., et al. 2007, ApJS, 171, 61  

\bibitem{} Heckman, T. M. 1980, A\&A, 87, 152

\bibitem{} Heisler, C.A., Lumsden, S.L., Bailey, J.A., 1997, Nature, 385, 700

\bibitem{} Hill, J. E., Burrows, D. N., Nousek, J. A., et al. 2004, Proc. SPIE, 5165, 217

\bibitem{Ho93} Ho, L.C., Filippenko, A.V., \& Sargent, W.L.W. 1993, ApJ, 
	417, 63

\bibitem{Ho97} Ho, L.C., Filippenko, A.V., \& Sargent, W.L.W. 1997, ApJS,
       112, 315

\bibitem{} Horne, K. 1986, PASP, 98, 609



\bibitem{} Jones, D.H., Saunders, W., Colless, M., et al. 2004, MNRAS, 355, 747

\bibitem{} Jones, D.H., Saunders, W., Read, M., Colless, M. 2005, PASA, 22, 277

\bibitem{} Kaspi, S., Smith, P.S., Netzer, H., et al. 2000, ApJ, 533, 631

\bibitem{} Kauffmann, G., Heckman, T.M., Tremonti, C., et al. 2003, 
	MNRAS, 346, 1055

	


\bibitem{} Landi, R., Bassani, L., Dean, A.J., et al. 2009, MNRAS, 392, 630

\bibitem{} Levenson, N.A.,  Weaver, K.A., Heckman, T.M., et al. 2005, ApJ, 618, 167


\bibitem{} Maiolino. R. Marconi, A., Salvati, M., et al. 2001, A\&A, 365, 28

\bibitem{} Malizia, A., Landi, R., Bassani, L., et al. 2007, ApJ, 668, 81



\bibitem{} Masetti, N., Palazzi, E., Bassani, L., et al. 2004, A\&A, 426, L41

\bibitem{} Masetti, N., Bassani, L., Bazzano, A., et al. 2006a, A\&A, 455, 
	11 

\bibitem{} Masetti, N., Morelli, L., Palazzi, E., et al. 2006b, A\&A, 459, 
	21 

\bibitem{} Masetti, N., Mason, E., Morelli, L., et al. 2008, A\&A, 482, 
	113 

\bibitem{} Masetti, N., Parisi, P., Palazzi, E., et al. 2009, A\&A, 495, 121

\bibitem{} Masetti, N., Parisi, P., Palazzi, E., et al. 2010, A\&A, 519, 96

\bibitem{} Masetti, N., Parisi, P., Jim\'enez-Bail\'on, E., et al. 2012, A\&A, 538, 123

\bibitem{} 	Massaro, E., Giommi, P., Leto, C., et al. 2009, A\&A, 495, 691 



\bibitem{} Monet, D.G., Levine, S.E., Canzian, B., et al. 2003, AJ, 125, 984  

\bibitem{} Moretti, A., Campana, S., Tagliaferri, G., et al. 2004, SPIE Proc., 5165, 232

\bibitem{} Mushotzky, R.F., Done, C., Pounds, K.A., 1993, ARA\&A, 31, 717


\bibitem{} Noguchi, K., Terashima Y. \& Awaki, H., 2009, ApJ, 705, 454

\bibitem{} Ojha, R., Fey, A.L., Johnston, K.J., et al. 2004, AJ, 127, 3609   

\bibitem{} Osterbrock, D.E., \& Pogge, R.W. 1985, ApJ, 297, 166

\bibitem{} Osterbrock, D.E. 1989, Astrophysics of Gaseous Nebulae and
       Active Galactic Nuclei (Mill Valley: Univ. Science Books)



\bibitem{} Panessa, F., de Rosa, A., Bassani, L., et al. 2011, MNRAS, 417, 2426

\bibitem{} Parisi, P., Masetti, N., Jim\'enez-Bail\'on, E., et al. 2009, A\&A, 507, 1345 

\bibitem{} Patterson, J. \& Raymond, J.C., 1985, ApJ, 292, 535

\bibitem{} Phillips, M.M., Charles, P.A., Baldwin, J.A., 1983, ApJ, 266, 485

\bibitem{} Predehl \& Schmitt, 1995, A\&A, 293, 889

\bibitem{} Prugniel, P., 2005, The Hyperleda Catalogue, http://leda.univ-lyon1.fr/








\bibitem{} 	Radivich, M.M. \& Kraus, J.D., 1971, AJ, 76, 683

\bibitem{} Ricci, R., Prandoni, I., Gruppioni, C., et al. 2004, 415, 549 

\bibitem{} Rieke \& Lebofski, 1985, ApJ, 288, 618




\bibitem{} Schlegel, D.J., Finkbeiner, D.P., \& Davis, M. 1998, ApJ, 500,
       525	


\bibitem{} Skrutskie, M.F., Cutri, R.M., Stiening, R., et al. 2006, AJ, 131, 1163 


\bibitem{} Struder L., Briel U., Dennerl K. et al. 2001, A\&A, 365L, 18



\bibitem{} Thuan, T.X, 1984, ApJ, 281, 126

\bibitem{} Trippe, M.L., Crenshaw, D.M., Deo, R.P., et al. 2010, ApJ. 725, 1749

\bibitem{} Ubertini, P., Lebrun, F., Di Cocco, G., et al. 2003, A\&A,
        411, L131



\bibitem{} 	Vaceli, M.S., Viegas, S.M., Gruenwald, R., et al. 1997, AJ, 114, 1345

\bibitem{} van Driel, W., Marcum, P., Gallagher, J. S., et al. 2001, A\&A, 378, 370

\bibitem{} Veilleux, S., \& Osterbrock, D.E. 1987, ApJS, 63, 295

\bibitem{} Veilleux, S., Godrich, R.W., Hill, G.J., 1997, ApJ, 477, 631

\bibitem{} Veron-Cetty, M. P. \& Veron, P., 2010, A\&A, 518, 10

\bibitem{} Voges, W., Aschenbach, B., Boller, T., et al. 1999, A\&A, 349, 389

\bibitem{} Warner, B. 1995, Cataclysmic variable stars (Cambridge: Cambridge University Press)

\bibitem{} Watson, M.G., Schr\"{o}der, A.C., Fyfe, D., et al. 2009, A\&A, 493, 339


\bibitem{} Winkler, H. 1992, MNRAS, 257, 677

\bibitem{} Winkler, C., Courvoisier, T.J.-L., Di Cocco, G., et al. 2003,
       A\&A, 411, L1


\bibitem{} Winter, L.M., Mushotzky, R.F., Reynolds, C.S., et al. 2009, ApJ, 690, 1322 


\bibitem{} Winter, L.M. \& Taylor, T., 2011, AAS, 21732606W

\bibitem{} Wright, E.L. 2006, PASP, 118, 1711

\bibitem{} Wu, X.-B., Wang, R., Kong, M.Z., Liu, F.K., \& Han, J.L. 2004,
       A\&A, 424, 793



\end{thebibliography}
\end{document}